# New Geometrical Approach to Superstrings

Alexander Belopolsky*
University of North Carolina
at Chapel Hill
E-mail: belopols@physics.unc.edu

April 1, 1997

## Abstract

We present a new geometrical approach to superstrings based on the geometrical theory of integration on supermanifolds. This approach provides an effective way to calculate multi-loop superstring amplitudes for arbitrary backgrounds. It makes possible to calculate amplitudes for the physical states defined as BRST cohomology classes using arbitrary representatives. Since the new formalism does not rely on the presence of primary representatives for the physical states it is particularly valuable for analyzing the discrete states for which no primary representatives are available. We show that the discrete states provide information about symmetries of the background including odd symmetries which mix Bose and Fermi states. The dilaton is an example of a non-discrete state which cannot be covariantly represented by a primary vertex. The new formalism allows to prove the dilaton theorem by a direct calculation.

## 1 Introduction and summary

Arguably, the most efficient method to describe string perturbation theory near a particular background is that of operator formalism combined with BRST quantization [1, 2]. In this approach a background can be described as an abstract conformal field theory (CFT) which has to be combined with a ghost system. Given a description of the background as a CFT one can calculate the spectrum of physical states and scattering amplitudes in a uniform manner. Since one can use the same formulae for different backgrounds, it becomes feasible to control the variation of the theory under the change of the background.

One of the predictions of the BRST approach is the appearance of the discrete states, or states which satisfy the physical state condition only for discrete values of momentum (in a Lorenz invariant theory discrete states may only appear at zero momentum). Although these states do not correspond to dynamical fields, the analysis of scattering amplitudes involving these fields can provide valuable information about the background. Such amplitudes are difficult to analyze mainly due to the fact that discrete states cannot be described by primary vertex operators.

*Supported in part by the US Department of Energy under Grant DE-FG 05-35ER40219/Task A.

Operator formalism for the bosonic string theory allows one to calculate amplitudes with the scattering states described as BRST classes. In short, given a set of BRST cohomology classes, the formalism allows one to construct a de Rham cohomology class of top degree forms on the moduli space. The resulting cohomology class can be integrated over the moduli space producing the amplitude. It is important that this formalism does not dictate what kind of representative for a BRST class one has to use. Primary representatives, when they exist, may simplify calculations but they are no longer a prerequisite.

Clearly the bosonic string operator formalism relies heavily on the theory of integration of differential forms. This makes it not trivial to apply the operator formalism to superstrings. The generalization of the main formula of the operator formalism, which gives the amplitude in terms of BRST cohomology classes of the scattering states was recently proposed by the author [3]. The present work contains a more elaborate exposition of the formalism presented in [3] and its applications.

We start in section 2 with a review of the theory of differential forms and geometric integration theory on supermanifolds. This review is based on the works of Bernstein and Leites [4,5], Baranov and Schwarz [6], and Voronov and Zorich [7,8,9,10,11] together with some new results by the author. We generalize differential forms of Voronov and Zorich by allowing them to have poles in the target space. This generalization will allow us to define operations analogous to the inner and exterior products which change the odd degree of a differential form. We will also be able to construct an analog the picture changing operator of the superstring BRST complex for the de Rham complex of a supermanifold.

In section 3 we will use the picture changing operator to analyze the geometry of superconformal surfaces which play the central role in superstring perturbation theory.

In section 4 we will review the construction of the BRST complex for the superstring with an emphasis on its geometric meaning. We will show that instead of the usual ghost number grading one has to use two gradings similar to even and odd dimension of the differential forms on the supermanifolds. The old ghost number can be recovered as the difference of these two new numbers.

In section 5 we define a notion of the superstring background. This is done in the most general setting when a background appears as an abstract superconformal field theory which may or may not correspond to a sigma-model on a Calabi-Yau manifold.

We conclude in section 6 by presenting several applications which illustrate the power of the geometric approach. The first application (section 6.1) is a geometrical proof of the dilaton theorem which is obtained by a straightforward evaluation of the formula for the string amplitudes. The dilaton theorem is an old result in string theory which shows that the vacuum expectation value of the dilaton field determines the string coupling constant (see [12,13]). Our geometric proof is very similar to the operator formalism approach to the dilaton theorem used in [14,15,16]. In particular the proof of the dilaton theorem for the heterotic string given in [15] is technically very similar to the one presented here. Another application is the analysis of the symmetries of the string backgrounds. The new formalism allows us to determine the global symmetries of the background by solving a cohomological problem. The resulting symmetries are exact to all orders in perturbation theory. As a particular example we show how the generators of the super-translation invariance can be obtained in the case of the flat $D = 10$ background.

## 2 Differential forms on supermanifolds

In this section we introduce the theory of differential forms on supermanifolds on which our approach to the superstrings will be based. We will start with a review of pseudodifferential and differential $r|s$-forms on the supermanifolds. In addition to the Cartan operators, the differential, the inner and exterior products, that change only the even degree, we will introduce operators that change odd degree of $r|s$-forms and define *picture changing operators*. As a marginal result we will find the super analog of the Plücker mapping.

### 2.1 Pseudodifferential forms

A common misconception in the theory of supermanifold is that differential forms have nothing to do with the theory of integration. Indeed, a naïve generalization of differential forms to the case of a supermanifold with even coordinates $x_a$ and odd coordinates $\xi_\alpha$ lead to polynomial expressions in *odd* symbols $\mathrm{d}x_a$ and *even* symbols $\mathrm{d}\xi_\alpha$ with coefficients being functions of $x_a$ and $\xi_\alpha$. We will see shortly that such objects are not suitable for integration [17].

In the pure even case the degree of a differential form can only be less or equal than the dimension of the manifold and the forms of the top degree transform as measures under smooth, orientation preserving coordinate transformations. This allows one to integrate the forms of the top degree over the oriented manifolds and forms of degree $k$ over the oriented $k$-dimensional subspaces. On a supermanifold, on the other hand, forms may have any degree due to the presence of commuting $\mathrm{d}\xi_\alpha$ and none of them transforms as a Berezinian measure.

One possible generalization of the differential forms which yields objects that can be integrated over supermanifolds was first suggested by Bernstein and Leites [4]. *Pseudodifferential forms* of Bernstein and Leites are defined as arbitrary *generalized* functions or distributions on $\hat{M} = \Pi TM$, where $M$ is the manifold and $\Pi$ indicates that the parity was changed to the opposite in the fibers of the tangent bundle.

Let $x = \{x^A\}$, where $A = 1, \ldots, m|n$ is a generalized index assuming $m$ even and $n$ odd values, be a set of coordinates on an $m|n$-dimensional manifold $M$. Naturally, $\mathrm{d}x = \{\mathrm{d}x^A\}$ provide a complete set of coordinates on the fibers of $\hat{M}$. Therefore, in the domain of definition of the coordinates $x$, one can write a pseudoform as $\omega(x, \mathrm{d}x)$ and define its integral as follows

$$\int_M \omega = \int_{\hat{M}} \mathcal{D}(x)\mathcal{D}(\mathrm{d}x)\,\omega(x, \mathrm{d}x). \tag{2.1.1}$$

Recall that if we make a change of coordinates from $x$ to $y$, the Berezin measure $\mathcal{D}(x)$ would transform into

$$\mathcal{D}(y) = \mathrm{sign}(\det J_{\bar{0}\bar{0}})\,\mathrm{Ber}\, J\, \mathcal{D}(x), \tag{2.1.2}$$

where $J = \partial y/\partial x$ is the Jacobi matrix and $J_{\bar{0}\bar{0}}$ is its even-even part. By definition the Jacobi matrix for the $\mathrm{d}x$ coordinates is $\Pi J$. Since

$$\mathrm{Ber}\,\Pi J = (\mathrm{Ber}\, J)^{-1} \text{ and } (\Pi J)_{\bar{0}\bar{0}} = J_{\bar{1}\bar{1}}, \tag{2.1.3}$$

the total measure in eq. (2.1.1) will only acquire a sign factor after the change of coordinates

$$\mathcal{D}(y)\mathcal{D}(\mathrm{d}y) = \mathrm{sign}(\det J_{\bar{0}\bar{0}})\,\mathrm{sign}(\det J_{\bar{1}\bar{1}})\mathcal{D}(x)\mathcal{D}(\mathrm{d}x). \tag{2.1.4}$$

Therefore, if we only allow coordinate transformations that preserve so-called $(--)$ orientation, or such that $\det J_{\bar{0}\bar{0}} \det J_{\bar{1}\bar{1}} > 0$, the measure will be invariant. In other words, for a super-manifold equipped with a $(--)$ orientation the integral of a pseudoform is unambiguously defined by eq. (2.1.1) (see [11, p. 74]).

Note that integral over $\mathrm{d}x$ in eq. (2.1.1) involves unbounded integrals in the even directions of the fibers of $\Pi TM$. It is necessary to require that pseudoforms vanish sufficiently fast in these directions. This explains why polynomial pseudoforms cannot be integrated.

Pseudoforms can also be integrated over submanifolds $N \subset M$ of arbitrary dimension by reducing the correspondent functions to $\Pi TN \subset \Pi TM$. With generalized functions the reduction is not always possible, for example a delta function supported at a point on a plane cannot be reduced to a generalized function on a line which contains this point.

The de Rham differential $\mathbf{d}$ is simply an odd vector field on $\Pi TM$ which can be written in coordinates as

$$\mathbf{d} = \sum_{A=1}^{m|n} \mathrm{d}x^A \frac{\partial}{\partial x^A}. \tag{2.1.5}$$

The property $\mathbf{d}^2 = 0$ follows trivially from the definition. Stokes theorem,

$$\int_N \mathbf{d}\omega = \int_{\partial N} \omega \tag{2.1.6}$$

is valid for pseudoforms. The Lie derivative and inner product are generalized easily to the case of pseudoforms; in coordinates,

$$\mathcal{L}_v = \sum_{A=1}^{m|n} v^A \frac{\partial}{\partial x^A} - \mathrm{d}x^A \frac{\partial v^B}{\partial x^A} \frac{\partial}{\partial (\mathrm{d}x^B)}, \tag{2.1.7}$$

$$\mathbf{i}_v = \sum_{A=1}^{m|n} v^A \frac{\partial}{\partial (\mathrm{d}x^A)}. \tag{2.1.8}$$

So far all the operators, $\mathbf{d}$, $\mathcal{L}_v$ and $\mathbf{i}_v$ appear as vector fields on $\Pi TM$ and they are related by the *homotopy identity*.

$$\mathcal{L}_v = [\mathbf{d}, \mathbf{i}_v], \tag{2.1.9}$$

where $[\cdot, \cdot]$ is the supercommutator of the differential operators, or equivalently, the super Lie bracket of the vector fields. External product of two pseudoforms is just the ordinary product of functions on $\Pi TM$ and as such, it is not always defined because pseudoforms are generalized functions. However, the product of a pseudoform with a linear function on $\Pi TM$ is always defined and thus we can introduce the exterior product operator $\mathbf{e}_\alpha$ for every 1-form $\alpha$

$$\mathbf{e}_\alpha \omega(x, \mathrm{d}x) = \alpha(x, \mathrm{d}x)\, \omega(x, \mathrm{d}x). \tag{2.1.10}$$

Operators $\mathbf{i}_v$ and $\mathbf{e}_\alpha$ satisfy the usual Clifford algebra commutation relations

$$[\mathbf{e}_\alpha, \mathbf{e}_\beta] = [\mathbf{i}_u, \mathbf{i}_v] = 0, \quad [\mathbf{i}_v, \mathbf{e}_\alpha] = \alpha(v). \tag{2.1.11}$$

The parity of these operators is given by

$$[\mathbf{e}_\alpha] = [\alpha] + \bar{1}, \quad [\mathbf{i}_v] = [v] + \bar{1}. \tag{2.1.12}$$

Due to the extremely simple form of the basic operations, pseudoforms are very convenient for calculations. Unfortunately, pseudoforms are not suitable for the superstring applications since one has to be able to integrate over supermanifolds with unoriented odd directions. Nevertheless, it will be possible to preserve the simplicity of the Cartan calculus for the pseudoforms by interpreting formulae (2.1.5), (2.1.7) and (2.1.8) in a formal manner. The corresponding formalism will be described in the following sections.

## 2.2 Differential $r|s$-forms

A complete theory of differential forms which have both even and odd degree was developed by Th. Voronov and A. Zorich (see ref. [11] and references therein) who generalized A. Schwarz's theory of densities on supermanifolds [18]. Densities on supermanifolds can be best described as (generalized) Lagrangians. A Lagrangian of degree $r|s$ is a function of a point on the supermanifold and a set of $r$ even and $s$ odd tangent vectors at this point. One can think of them as Lagrangians describing the dynamics of $(r|s-1)$-dimensional extended objects propagating through the supermanifold. Let $x^A$, be coordinates on $M$ and $t^F$, $F = 1, \ldots, r|s$ be coordinates on another manifold $N$ and $x(t)$ be a mapping $N \to M$. We will use the notation $\dot{x}$ for the matrix $\{\partial x^A / \partial t^F\}$ or for the set of $r$ even and $s$ odd (column) vectors at any point $x$ in $M$ depending on the context. With this notation we can write the action associated with the Lagrangian $L(x, \dot{x})$ simply as

$$S[N] = \int_N \mathcal{D}(t) \, L(x(t), \dot{x}(t)). \tag{2.2.1}$$

A Lagrangian is called *covariant* if its action does not depend on the parameterization of $N$. According to the formula for the change of variables in the Berezin integral, being covariant is equivalent to

$$L(x, J\dot{x}) = \operatorname{sign}(\det J_{\bar{0}\bar{0}}) \operatorname{Ber} J \, L(x, \dot{x}), \quad \text{for any } J \in GL(r|s). \tag{2.2.2}$$

Covariant Lagrangians are also known as densities on supermanifolds [18].

Differential $r|s$-forms are defined as Lagrangians of degree $r|s$ that satisfy two conditions. First,

$$L(x, J\dot{x}) = \operatorname{Ber} J \, L(x, \dot{x}), \quad \text{for any } J \in GL(r|s), \tag{2.2.3}$$

which is equivalent to the covariance condition if $\det J_{\bar{0}\bar{0}} > 0$ and therefore makes $r|s$-forms suitable for integration over manifolds with $(+-)$ orientation [11, p. 56]. Second, it is required that the variations of the corresponding actions do not contain second derivatives. Explicitly,

$$\frac{\delta S}{\delta x^A} = \frac{\delta L}{\delta x^A} - (-1)^{[A][F]} \frac{1}{2} \ddot{x}^B_{FG} \left( \frac{\partial^2 L}{\partial \dot{x}^A_F \partial \dot{x}^B_G} + (-1)^{[F][G]+[B]([F]+[G])} \frac{\partial^2 L}{\partial \dot{x}^A_G \partial \dot{x}^B_F} \right),$$

where

$$\frac{\delta L}{\delta x^A} \stackrel{\text{def}}{=} \frac{\partial L}{\partial x^A} - (-1)^{[A][F]} \dot{x}^B_F \frac{\partial^2 L}{\partial x^B \partial \dot{x}^A_F}. \tag{2.2.4}$$

In other words, a differential $r|s$-form must satisfy the following *basic equations*

$$\frac{\partial^2 L}{\partial \dot{x}^A_F \partial \dot{x}^B_G} + (-1)^{[F][G]+[B]([F]+[G])} \frac{\partial^2 L}{\partial \dot{x}^A_G \partial \dot{x}^B_F} = 0. \tag{2.2.5}$$

Voronov and Zorich [11, p. 57] require that differential forms be defined everywhere on $W_{r|s}(M)$, the bundle of $r|s$-tuples of tangent vectors $F$ such that rank $F_{\bar{1}\bar{1}} = s$. In the following we will relax this condition and allow forms that have poles on $W_{r|s}(M)$. The space of $r|s$-forms will be denoted by $\Omega^{r|s}(M)$.

De Rham differential $\mathbf{d}$ maps $r|s$ forms to $r+1|s$ forms and is given by

$$\mathbf{d}L = (-1)^r \dot{x}^A_{r+1} \frac{\delta L}{\delta x^A}. \tag{2.2.6}$$

The usual properties, $\mathbf{d}^2 = 0$ and the Stokes theorem remain valid for the forms on supermanifolds. The inner product $\mathbf{i}_v$ is defined as a substitution replacing the first vector in $\dot{x}$. Since $r|s$-forms are multilinear and skew-symmetric in even vectors, the following explicit formula for the inner product holds

$$\mathbf{i}_v L = (-1)^{r-1} v^A \frac{\partial L}{\partial \dot{x}_r^A} \tag{2.2.7}$$

Lie derivative along a vector field is defined as usual, explicitly

$$\mathcal{L}_v L = v^A \frac{\partial L}{\partial x^A} + (-1)^{[F][v]} \dot{x}_F^B \frac{\partial v^A}{\partial x^B} \frac{\partial L}{\partial \dot{x}_F^A} \tag{2.2.8}$$

Exterior product operator $\mathbf{e}_\alpha$, which maps $\Omega^{r|s}(M)$ to $\Omega^{r+1|s}(M)$ is given by

$$\mathbf{e}_\alpha L = (-1)^r \left[ \alpha(\dot{x}_{r+1}) - (-1)^{[\alpha][F]} \alpha(\dot{x}_F) \dot{x}_{r+1}^A \frac{\partial}{\partial \dot{x}_F^B} \right] L. \tag{2.2.9}$$

One can see that the explicit formulae expressing the Cartan operators for the $r|s$-forms are much more complicated than the corresponding formulae for the pseudoforms.

## 2.3 Baranov-Schwarz transformation

Differential $r|s$-forms and pseudoforms are super analogs of $k$-forms and inhomogeneous forms (formal linear combinations of differential forms of different degree) in pure even geometry. To make the analogy complete we have to present the projection of pseudoforms onto $r|s$-forms for any given $r$ and $s$. The integral of a pseudoform $\omega(x, \mathrm{d}x)$ over a submanifold $N$ can be written as

$$\int_N \omega = \int_{\Pi TN} \mathcal{D}(t) \mathcal{D}(\mathrm{d}t) \omega(x(t), \mathrm{d}x(t)) \tag{2.3.1}$$

Expanding the differential $\mathrm{d}x(t) = \sum_{F=1}^{r|s} \mathrm{d}t^F \dot{x}_F^A$ and taking the integral with respect to $\mathrm{d}t$ we obtain

$$\int_N \omega = \int_N \mathcal{D}(t) L_\omega^{r|s}(x(t), \dot{x}(t)) = \int_N L_\omega^{r|s}, \tag{2.3.2}$$

where

$$L_\omega^{r|s}(x, \dot{x}) \stackrel{\mathrm{def}}{=} \int_{\Pi T_x N} \mathcal{D}(\mathrm{d}t) \omega\Big(x, \sum_{F=1}^{r|s} \mathrm{d}t^F \dot{x}_F^A\Big). \tag{2.3.3}$$

The resulting projection,

$$\lambda^{r|s} : \omega \mapsto L_\omega^{r|s}, \tag{2.3.4}$$

is known as the Baranov-Schwarz transformation [6]. The resulting Lagrangian satisfies the basic equations (2.2.5), but its transformation property differs by a sign factor both from eq. (2.2.3) and eq. (2.2.2)

$$L_\omega^{r|s}(x, J\dot{x}) = \text{sign}(\det J_{\bar{1}\bar{1}}) \, \text{Ber}\, J \, L_\omega^{r|s}(x, \dot{x}). \tag{2.3.5}$$

Such Lagrangians are called $r|s$-forms of type II. Note that on the supermanifolds with a $(++)$ orientation, or the orientation defined separately in both even and odd direction there is no difference between type I and type II forms.

All the operations defined so far, de Rham differential $\mathbf{d}$, Lie derivative $\mathcal{L}_v$, and inner and exterior products, $\mathbf{i}_v$ and $\mathbf{e}_\alpha$ are consistent with respect to the Baranov-Schwarz transformation.

## 2.4 Plücker mapping and the de Rham complex

Consider $r$ even and $s$ odd covectors $\alpha^F$, $F = 1, \ldots, r|s$ forming a non-degenerate co-frame. We can define a pseudoform

$$\omega(x, \mathrm{d}x) = \prod_{F=1}^{r|s} \delta(\alpha^F(x, \mathrm{d}x)), \tag{2.4.1}$$

which is simply a $\delta$-function supported on the zero-space of $\alpha^F$. After an application of the Baranov-Schwarz transformation, to this pseudoform, we obtain a type II form

$$L_\omega^{II}(x, \dot{x}) = \text{sign}(\det \|\alpha^F(x, \dot{x}_G)\|_{\bar{1}\bar{1}}) \, \text{Ber}\|\alpha^F(x, \dot{x}_G)\|. \tag{2.4.2}$$

We will call these $r|s$-forms Plücker forms, since their construction provides the super analog of the Plücker map

$$P\ell: \, Gr(m - r|n - s, \mathbb{R}^{m|n}) \to P(\Lambda^{r|s}(\mathbb{R}^{m|n})), \tag{2.4.3}$$

where $\Lambda^{r|s}(\mathbb{R}^{m|n})$ is the space of $r|s$-forms "at a point". The covectors $\alpha^F$ will be called generators of the Plücker form.

By analogy with eq. (2.4.2) we can define a Plücker form of type I as

$$L_\omega(x, \dot{x}) = \text{Ber}\|\alpha^F(x, \dot{x}_G)\|. \tag{2.4.4}$$

Although $L_\omega$ is not directly related to the pseudoforms $\omega$, it will be convenient to consider eq. (2.4.1) as an alternative way to represent an $r|s$-form. Of course, the direct interpretation of $\delta$-functions as generalized functions will then be impossible.

If two Plücker forms have linearly independent generators, they can be multiplied and produce another Plücker form which is generated by the combination of the two sets of generators. In terms of the Grassmanian the product is equivalent to the intersection of subspaces. Clearly, when the product is defined, its dimension is the sum of the dimensions of the multipliers.

Given a closed base Plücker $r|s$-form $L$ on $M$ we define the complex $\Omega_L^*(M)$ which is generated from $L$ by the action of $\mathbf{i}_v$ and $\mathcal{L}_v$ for all vector fields $v$ and $\mathbf{e}_\alpha$ for all one-forms $\alpha$. All the forms in $\Omega_L^*(M)$ will have the same odd degree as the base form $L$ since

$$\deg(\mathbf{e}_\alpha) = 1|0, \quad \deg(\mathbf{i}_v) = -1|0, \quad \text{and} \quad \deg(\mathcal{L}_v) = 0|0. \tag{2.4.5}$$

In the two limiting cases $\deg L = 0$ and $\deg L = \dim M$, $\Omega_L^*(M)$ reproduces two well known complexes: one of polynomial differential forms and the other of integral forms on $M$ (see [19, 4, 5])

$$\Omega_{f(x)}^*(M) = \Omega_{\text{pol}}^*(M), \quad \Omega_{f(x)\,\text{Ber}\,\dot x}^*(M) = \Sigma^*(M). \tag{2.4.6}$$

The first complex consists of forms of positive even degree and the second will have the even degree bounded from above by the even dimension of the manifold $M$. For any other choice of the base form $L$, the complex $\Omega_{\omega_0}^*(M)$ is infinite in both directions. Forms with even degree less than zero cannot be described by Lagrangians, but they have to be included in the de Rham complex anyway [20].

The de Rham differential maps $\Omega_L^n(M)$ to $\Omega_L^{n+1}(M)$ and turns $\Omega_L^*(M)$ into a differential complex, defining yet another cohomology on the supermanifold.

Although we will be only working with the type I differential forms, it will be convenient to formally write them as pseudoforms by representing a Plücker form by a product of delta functions, inner product as a derivative and the exterior product by a multiplication. When they appear in the notation for the type I forms, delta functions will have some unusual properties such as

$$\delta(\mathrm{d}\xi)\delta(\mathrm{d}\eta) = -\delta(\mathrm{d}\eta)\delta(\mathrm{d}\xi) \quad \text{and} \quad \delta(-\mathrm{d}\xi) = -\delta(\mathrm{d}\xi), \tag{2.4.7}$$

which follow from the transformation law eq. (2.2.3). On the other hand they will in many aspects behave just as usual delta-functions, for example $\mathrm{d}\xi\,\delta(\mathrm{d}\xi)$ will still be zero. Note that this pseudoform-like notation will only work for the differential forms from a de Rham complex based on a Plücker form. It is possible that there are forms that cannot be represented in this way but examples are hard to find.

## 2.5 Picture changing operator

Clifford algebra operators $\mathbf{i}_v$ and $\mathbf{e}_\alpha$ change only even degree of differential forms even when $v$ or $\alpha$ is odd. In order to change the odd degree one has to use delta functions of these operators. Let $v$ and $\alpha$ be odd vector field and odd 1-form respectively. We define operators $\delta(\mathbf{i}_v)$ and $\delta(\mathbf{e}_v)$ acting on pseudoforms as follows

$$[\delta(\mathbf{i}_v)\omega](x, \mathrm{d}x) = \int_{-\infty}^{\infty} \mathrm{d}t\,\omega(x, \mathrm{d}x + t\,v) \tag{2.5.1}$$

and

$$[\delta(\mathbf{e}_\alpha)\omega](x, \mathrm{d}x) = \delta(\alpha(x, \mathrm{d}x))\,\omega(x, \mathrm{d}x). \tag{2.5.2}$$

The first of the two definitions can be justified using the integral form of the delta function. Applying the Baranov-Schwarz transformation to eq. (2.5.1), we find that $\delta(\mathbf{i}_v)$ is equivalent to a substitution of $v$ into the Lagrangian replacing an odd vector

$$[\delta(\mathbf{i}_v)L](x; \dot{x}) = L(x; \dot{x}, v), \tag{2.5.3}$$

and $\delta(\mathbf{e}_\alpha)$ is given by

$$[\delta(\mathbf{e}_\alpha)L](x; \dot{x}, v) = \frac{1}{\alpha(v)} L\left(x, \dot{x} - \frac{\alpha(\dot{x})v}{\alpha(v)}\right), \tag{2.5.4}$$

where $v$ is the vector on the $s+1$-st odd place. Delta function operators have the following properties that up-to a sign one would naturally expect from a delta function

$$\mathbf{i}_v\,\delta(\mathbf{i}_v) = \mathbf{e}_\alpha\,\delta(\mathbf{e}_\alpha) = 0, \tag{2.5.5}$$

if $\lambda$ is an arbitrary non-vanishing[1] function then

$$\delta(\mathbf{e}_{\lambda\cdot\alpha}) = \frac{1}{\lambda}\delta(\mathbf{e}_\alpha), \quad \delta(\mathbf{i}_{\lambda\cdot v}) = \frac{1}{\lambda}\delta(\mathbf{i}_v), \tag{2.5.6}$$

and

$$\delta(a\,x + b\,y)\,\delta(c\,x + d\,y) = \frac{1}{\det\begin{pmatrix} a & b \\ c & d \end{pmatrix}} \delta(x)\,\delta(y), \tag{2.5.7}$$

In particular $\delta(y)\,\delta(x) = -\delta(x)\,\delta(y)$ i.e., the product is anticommutative and

$$\delta(\mathbf{i}_u)\delta(\mathbf{i}_v) = -\delta(\mathbf{i}_v)\delta(\mathbf{i}_u), \quad \delta(\mathbf{e}_\alpha)\delta(\mathbf{e}_\beta) = -\delta(\mathbf{e}_\beta)\delta(\mathbf{e}_\alpha). \tag{2.5.8}$$

An important operation which lowers the odd dimension of a differential form is "partial integration" along a non-vanishing odd vector field. An odd vector field $\nu$ generates *odd exponential map*, which we define as

$$\exp^\nu: \mathbb{R}^{0|1} \times M \to M, \quad x^A \mapsto \exp\left(\tau\nu^B \frac{\partial}{\partial x^B}\right) x^A = x^A + \tau\,\nu^A, \tag{2.5.9}$$

where $x^A$ are coordinate functions on $M$ and $\tau$ is the coordinate on $\mathbb{R}^{0|1}$. Pseudoforms on $M$ can be pulled back to $\mathbb{R}^{0|1} \times M$ and then integrated over $\tau$. We define "picture-changing" operator[2] $\Gamma_\nu$ as follows

$$[\Gamma_\nu\,\omega](x, \mathrm{d}x) = \int \mathcal{D}(\tau)\mathcal{D}(\mathrm{d}\tau)\,\omega(x + \tau\,\nu, \mathrm{d}x + \mathrm{d}\tau\,\nu - \tau\,\mathrm{d}x^A\,\frac{\partial\nu}{\partial x^A}), \tag{2.5.10}$$

---

[1] In the context of supermanifolds "non-vanishing" means that the function is not zero when all it's odd arguments are set to zero.

[2] The name is borrowed from a similar operation on the BRST complex of the superstring [21]

where the second argument to $\omega$ is just $\mathrm{d}(x+\tau\nu)$. The integral over the odd parameter $\tau$ can be performed easily,

$$[\Gamma_\nu \omega](x,\mathrm{d}x) = \int \mathcal{D}(\mathrm{d}\tau) \left[\left(\nu^A \frac{\partial}{\partial x^A} - \mathrm{d}x^A \frac{\partial \nu^B}{\partial x^A} \frac{\partial}{\partial (\mathrm{d}x^B)}\right)\omega\right](x, \mathrm{d}x + \mathrm{d}\tau\,\nu)$$

$$= \int \mathcal{D}(\mathrm{d}\tau)\,[(\mathcal{L}_\nu + \mathrm{d}\tau\,\mathbf{i}_{\nu^2})\,\omega]\,(x, \mathrm{d}x + \mathrm{d}\tau\,\nu)$$

$$= [(\delta(\mathbf{i}_\nu)\mathcal{L}_\nu + \delta'(\mathbf{i}_\nu)\,\mathbf{i}_{\nu^2})\omega](x,\mathrm{d}x), \quad (2.5.11)$$

where

$$\nu^2 = \nu^A \frac{\partial \nu}{\partial x^A} = \frac{1}{2}\{\nu,\nu\}, \quad (2.5.12)$$

and

$$\delta'(\mathbf{i}_\nu)\omega(x,\mathrm{d}x) = \int\limits_{-\infty}^{\infty} \mathrm{d}t\, t\,\omega(x, \mathrm{d}x + t\,\nu). \quad (2.5.13)$$

At this point we can express the picture changing operator as follows

$$\Gamma_\nu = \delta(\mathbf{i}_\nu)\mathcal{L}_\nu + \delta'(\mathbf{i}_\nu)\,\mathbf{i}_{\nu^2} = \frac{1}{2}(\delta(\mathbf{i}_\nu)\mathcal{L}_\nu + \mathcal{L}_\nu\delta(\mathbf{i}_\nu)) \quad (2.5.14)$$

Picture changing operator can also be defined directly on differential $r|s$-forms transformation. We define

$$\Gamma_\nu : \Omega^{r|s}(M) \to \Omega^{r|s-1}(M), \quad (2.5.15)$$

and for $L(x;\dot{x}) \in \Omega^{r|s}$,

$$[\Gamma_\nu L](x;\dot{x}') = \int \mathcal{D}(\tau) L(x + \tau\nu(x);\ \dot{x}' + \tau\,[\nu(x),\dot{x}'],\ \nu(x+\tau\nu(x))), \quad (2.5.16)$$

where $\dot{x}'$ stands for a set of $r$ even and $s-1$ odd vectors.

The picture changing operator has the following important property: $\Gamma_\nu$ does not change when $\nu$ is multiplied by an even non-vanishing function,

$$\Gamma_{\lambda\cdot\nu} = \Gamma_\nu. \quad (2.5.17)$$

This property is rather natural since $\Gamma_\nu$ is an operation of integration over an odd submanifold generated by $\nu$ and vector fields that differ by a factor generate the same submanifold.

The most important property of the picture-changing operator is that it commutes with the de Rham differential:

$$[\mathbf{d},\Gamma_\nu] = 0. \quad (2.5.18)$$

This property can be established by a direct computation using explicit formulae, but it is useful to give a heuristic argument. Recall that $\mathcal{L}_\nu = [\mathbf{d}, \mathbf{i}_\nu]$ and thus formally

$$\Gamma_\nu = [\mathbf{d}, \Theta(\mathbf{i}_\nu)] = \delta(\mathbf{i}_\nu)\mathcal{L}_\nu + \delta'(\mathbf{i}_\nu)\,\mathbf{i}_{\nu^2}, \qquad (2.5.19)$$

where $\Theta$ is the Heavyside step function $\Theta'(x) = \delta(x)$ and

$$[\Theta(\mathbf{i}_\nu)\omega](x, \mathrm{d}x) = \int_{-\infty}^{\infty} \frac{\mathrm{d}t}{t}\,\omega(x, \mathrm{d}x + t\,\nu). \qquad (2.5.20)$$

The second term of $\Gamma_\nu$ appears in eq. (2.5.19) because $\mathcal{L}_\nu$ does not commute with $\mathbf{i}_\nu$,

$$[\mathcal{L}_\nu, \mathbf{i}_\nu] = \mathbf{i}_{\{\nu,\nu\}} = 2\,\mathbf{i}_{\nu^2}. \qquad (2.5.21)$$

It would be interesting to define higher order picture changing operators integrating over higher dimensional odd subspaces. It is expected[3] that these higher order operators would have the form

$$\Gamma^{(k)}_{\nu_1\nu_2\cdots\nu_k} = \Gamma_{\nu_1}\Gamma_{\nu_k}\cdots\Gamma_{\nu_k} + [\mathbf{d}, \cdots], \qquad (2.5.22)$$

but a complete construction is not available yet.

## 3 Superconformal surfaces

In this section we describe a special kind of dimension 2|2 supermanifolds known as superconformal surfaces. We will show that in many aspects superconformal surfaces resemble one-dimensional complex curves and that the theory of integration on them is much simpler than that on a general supermanifold. Superconformal surfaces are sometimes referred in the physics literature as super-Riemann surfaces. Following refs. [23, 24] we reserve the name "super-Riemann surface" or "SRS" for different objects. The distinction between an SRS and a superconformal surface is similar to the distinction between ordinary two-dimensional conformal surfaces and ordinary Riemann surfaces. Riemann surfaces come equipped with a metric while on conformal surfaces metrics is defined only up to a scalar factor. Superconformal surfaces were extensively studied by a number of authors [25, 26, 27, 28, 29, 30, 31, 32, 33, 23, 34, 24, 35, 36]. Here we will reproduce some of the results using the technique based on picture changing operators.

### 3.1 Differential geometry on a superconformal surface

A superconformal surface is defined as a dimension 1|1 complex supermanifold with a distinguished analytical odd vector field $D$ defined up to a multiplication

---
[3]The following conjecture is based on a similar result for the superstring BRST complex [22].

by a non-vanishing function. It is required that $D$ and $D^2 = \frac{1}{2}\{D,D\}$ form a basis of the complex tangent space at every point. One can show that in the neighborhood of any point on a superconformal manifold there exist a local coordinate system $\boldsymbol{z} = (z, \theta)$ such that

$$D \propto D_\theta = \frac{\partial}{\partial \theta} + \theta \frac{\partial}{\partial z}. \tag{3.1.1}$$

We will call such coordinate system superconformal and $\boldsymbol{z} = (z, \theta)$ superconformal coordinates. If $\boldsymbol{z}$ and $\tilde{\boldsymbol{z}}$ are two sets of superconformal coordinates defined on the same domain, then eq. (3.1.1) requires that $D_\theta \propto D_{\tilde{\theta}}$ or

$$D_\theta \tilde{z} = \tilde{\theta} D_\theta \tilde{\theta} \tag{3.1.2}$$

and

$$D_{\tilde{\theta}} = F(\boldsymbol{z}) \cdot D_\theta, \quad \text{where } F(\boldsymbol{z}) = D_\theta \tilde{\theta}. \tag{3.1.3}$$

Following [37], we define a *superconformal field* of dimension $(h, \bar{h})$ as an object given locally by a function $\Phi(\boldsymbol{z}, \bar{\boldsymbol{z}})$ and transforms under superconformal transformations according to

$$\tilde{\Phi}(\tilde{\boldsymbol{z}}, \tilde{\bar{\boldsymbol{z}}}) = F(\boldsymbol{z})^{-2h} \bar{F}(\bar{\boldsymbol{z}})^{-2\bar{h}} \Phi(\boldsymbol{z}, \bar{\boldsymbol{z}}). \tag{3.1.4}$$

Superconformal fields can be multiplied and the dimension of the product is the sum of the dimensions of the factors.

Superconformal fields of dimension $(-1, 0)$ and $(-1/2, 0)$ are closely related to superconformal vector fields. Establishing this relation, we assign the odd superconformal vector field

$$u = U(\boldsymbol{z}) D_\theta \tag{3.1.5}$$

to a $(-1/2, 0)$ superconformal field $U$ and the even superconformal vector field

$$v = V(\boldsymbol{z}) \frac{\partial}{\partial z} + \frac{1}{2} D_\theta V(\boldsymbol{z}) D_\theta \tag{3.1.6}$$

to a $(-1, 0)$ field $V$.

The resulting vector fields are related by $v = u^2 \stackrel{\text{def}}{=} \frac{1}{2}\{u, u\}$ when $V = U^2$. Vector fields corresponding to $(-1/2, 0)$ correspond to infinitesimal superconformal transformations. The simplest way to see that an infinitesimal transformation corresponding to the vector field of the form in eq. (3.1.6) preserves $D$ up to a factor and, therefore, is superconformal, is to calculate the Lie bracket

$$[v, D] = \left[V(\boldsymbol{z}) \frac{\partial}{\partial z} + \frac{1}{2} D_\theta V(\boldsymbol{z}) D_\theta, D_\theta\right] = -\frac{1}{2} \partial V \, D_\theta \propto D. \tag{3.1.7}$$

Superconformal fields of dimension $(0, 0)$ and $(1/2, 0)$ are related to differential forms. This relation can be established in two different pictures. To superconformal fields of dimension $(0,0)$ and $(1/2, 0)$ respectively, we assign 0|0-forms (functions) and 1|0-forms (polynomial one-forms) in picture zero and 0|1- and 1|1-forms in picture one. Explicit formulae in superconformal coordinates $\boldsymbol{z} = (z, \theta)$ are given in Table 1.

| Field | Lagrangian | Pseudoform |
|---|---|---|
| $\Phi_0$ | $L^{0\|0} = \Phi_0(\boldsymbol{z})$ | $\Phi_0(\boldsymbol{z})$ |
| $\Phi_0$ | $L^{0\|1} = \left(\dfrac{\theta}{\hat{\theta}} - \dfrac{\hat{z}}{\hat{\theta}^2}\right)\Phi_0(\boldsymbol{z})$ | $\left(\mathrm{d}z\,\delta'(\mathrm{d}\theta) + \theta\,\delta(\mathrm{d}\theta)\right)\Phi_0(\boldsymbol{z})$ |
| $\Phi_{1/2}$ | $L^{1\|0} = \dot{\theta}\,\Phi_{1/2}(\boldsymbol{z}) + (\dot{z} - \theta\,\dot{\theta})\,D_\theta\Phi_{1/2}(\boldsymbol{z})$ | $\mathrm{d}\theta\,\Phi_{1/2}(\boldsymbol{z}) + (\mathrm{d}z - \theta\,\mathrm{d}\theta)\,D_\theta\Phi_{1/2}(\boldsymbol{z})$ |
| $\Phi_{1/2}$ | $L^{1\|1} = \mathrm{Ber}\begin{pmatrix}\dot{z} & \dot{\theta} \\ \hat{z} & \hat{\theta}\end{pmatrix}\Phi_{1/2}(\boldsymbol{z})$ | $\mathrm{d}z\,\delta(\mathrm{d}\theta)\,\Phi_{1/2}(\boldsymbol{z})$ |

Table 1: Lagrangians and pseudoforms assigned to different superconformal fields. $\Phi_h$ stands for a superconformal field of dimension $(h, 0)$. Vector arguments of the Lagrangians are $(\dot{z}, \dot{\theta})$ for an even vector and $(\hat{z}, \hat{\theta})$ for an odd one.

Differential forms in the two pictures are related by the picture changing operator $\Gamma_D$. Recall that the picture changing operator $\Gamma_\nu$ does not change when $\nu$ is multiplied by a factor. This means that a superconformal surface $\Sigma$ has just enough structure for $\Gamma_D : \Omega^{r|s}\Sigma \to \Omega^{r|s-1}\Sigma$ to be well defined. Moreover, all four Lagrangians in Table 1 can be related using $D$

$$\begin{aligned} L^{0|0} &= \Gamma_D\, L^{0|1} & L^{0|0} &= \mathbf{i}_D\, L^{1|0} \\ L^{1|0} &= \Gamma_D\, L^{1|1} & L^{0|1} &= \mathbf{i}_D\, L^{1|1} \end{aligned} \tag{3.1.8}$$

For the future references we denote the factors that superconformal fields acquire in picture one as follows

$$\omega^{0|1} = \mathrm{d}z\,\delta'(\mathrm{d}\theta) + \theta\,\delta(\mathrm{d}\theta), \quad \text{and} \quad \omega^{1|1} = \mathrm{d}z\,\delta(\mathrm{d}\theta). \tag{3.1.9}$$

### 3.2 Contour integration on superconformal manifolds

Integrals of superconformal fields can be defined as the integrals of the corresponding differential forms. Therefore, superconformal fields of dimension $(1/2, 0)$ can be integrated either over $1|0$- or $1|1$-dimensional submanifolds depending on the picture and superconformal fields of dimension $(0, 0)$ can be evaluated at points or integrated over $0|1$-dimensional submanifolds.

We will call submanifolds of real dimension $1|0$ and $1|1$ respectively $1|0$- and $1|1$-contours and $0|1$-dimensional submanifolds will be referred to as $0|1$-points. Recall that at every point on a superconformal manifold there is a distinguished dimension $0|1$ subspace spanned by $D$. A $0|1$-point is called superconformal if it is tangent to $D$, similarly a $1|1$-contour is called superconformal if it is tangent to $D$ and $D^2$ and its boundary is either empty or consists of superconformal $0|1$-points. With every point one can associate a superconformal $0|1$-point and with every $1|0$-contour a superconformal $1|1$-contour by means of a map which we denote by $\Gamma_D^+$. In superconformal coordinates

$$\Gamma_D^+ : \begin{array}{rl} (z_0, \theta_0) & \to (z_0 + \tau\,\theta_0, \theta_0 + \tau) \\ (f(t), \psi(t)) & \to (f(t) + \tau\,\psi(t)\,\sqrt{f'(t)}, \\ & \quad \psi(t) + \tau\,\sqrt{f'(t) + \psi(t)\psi'(t)}) \end{array} \quad (3.2.1)$$

Notation $\Gamma_D^+$ is justified by the following property of this map:

$$\int_{\Gamma_D^+ N} L = \int_N \Gamma_D L, \quad (3.2.2)$$

where $N$ is either $0|0$- or $1|0$-dimensional and $L$ is respectively either $0|1$- or $1|1$-form. Therefore, if we consider taking an integral as a pairing between submanifolds and differential forms then $\Gamma_D^+$ is an adjoint operator to $\Gamma_D$.

Operator $D$ is closely related to the de Rham differential. Let $\Phi$ be a function on a superconformal manifold then $D\Phi$ is a dimension $(1/2, 0)$ superconformal field. One can show that

$$L_{D\Phi} = \mathbf{d} L_\Phi \quad (3.2.3)$$

is valid when Lagrangians are taken in the same picture. Taken in picture one, eq. (3.2.3) together with the Stokes theorem leads to the following simple rule [34]

$$\int_{P_1}^{P_2} \mathrm{d}\boldsymbol{z}\, D\Phi(\boldsymbol{z}) = \Phi(P_2) - \Phi(P_1), \quad (3.2.4)$$

where $P_1$ and $P_2$ are two points and the integral is taken over any $1|1$-contour whose boundary is $\Gamma_D^+(P_2) - \Gamma_D^+(P_1)$ and we have just introduced a new notation

$$\mathrm{d}\boldsymbol{z} = \mathrm{Ber}\begin{pmatrix} \dot{z} & \dot{\theta} \\ \hat{z} & \hat{\theta} \end{pmatrix} \quad (3.2.5)$$

Note that unlike in [34], the contour of integration is not required to be superconformal, only its boundary.

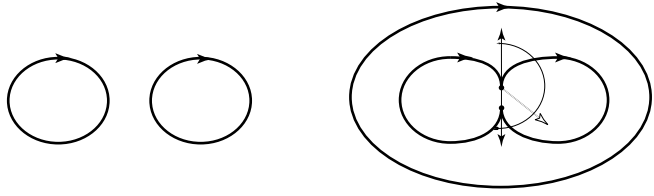

Figure 1: Deformation of superconformal contours.

Using the $\mathrm{d}z$ symbol we can write a $(h,0)$ superconformal field covariantly as

$$\phi = \phi(z,\bar z)(\mathrm{d}z)^{2h}. \tag{3.2.6}$$

Making the analogy with the pure even case even closer we will use the term $(2h)$-differential to refer to $(h,0)$ superconformal fields.

Every closed real 1|1-dimensional connected supermanifold is either $S^{1|1}$ (with a circle as carrier and non-trivial normal bundle), or its double-valued cover $S^1 \times \mathbb{R}^{0|1}$. Therefore, on a superconformal manifold there are two kinds of cycles. A cycle will be called Neveu-Schwarz (NS) (respectively Ramond (R)) if it is isomorphic to $S^{1|1}$ (respectively $S^1 \times \mathbb{R}^{0|1}$). We will use $t$ and $\tau$ for the coordinates on the cycles and identify $(t,\tau) = (t+2\pi, -\tau)$ for an NS-cycle and $(t,\tau) = (t+2\pi, \tau)$ for an R-cycle. A canonical example of a superconformal cycle is provided by

$$(t,\tau) \mapsto (z = e^{it},\, \theta = e^{it/2}\tau), \tag{3.2.7}$$

where $z$ and $\theta$ are superconformal coordinates. As usual, $\theta$ has to be multivalued in order for eq. (3.2.7) to describe an R-cycle. On the other hand, if we do not require an R-cycle to be superconformal then it can be easily described within a single coordinate patch, e.g. $(t,\tau) \mapsto (z = e^{it},\, \theta = \tau)$. An important lesson is that although an R-cycle can be drawn anywhere on a superconformal manifold, a *superconformal* R-cycle has to wrap around a non-trivial cycle or a puncture on the reduced manifold. One can split and join superconformal 1|1-cycles by splitting and joining their corresponding 1|0-cycles. The integrals of closed 1|1-forms like $\mathrm{d}z\,\Phi(z)$ will not change.

It is interesting to see why a joint of two R-cycles gives an NS-cycle. At the first glance it is counter intuitive because NS-cycles are those with a twist while R-cycles are without. Let us show that if $\Gamma_D^+(\gamma_1)$ and $\Gamma_D^+(\gamma_2)$ are Ramond then $\Gamma_D^+(\gamma_1 + \gamma_2)$ is Neveu-Schwarz, where $\gamma_1 + \gamma_2$ is obtained by deformation from $\gamma_1$ and $\gamma_2$. Recall that the odd parameter $\tau$ of a 1|1-cycles obtained by $\Gamma_D^+(f(t),\psi(t))$ appears with a factor $\sqrt{f'(t)}$. As shown in Fig. 1, when we join two cycles, the resulting cycle has two singularities where $f'(t)$ changes sign and, therefore, $\sqrt{f'(t)}$ changes its phase by $\pi/2$. Together these two phase changes introduce a change of sign for $\tau$. Similarly we can conclude that NS+NS=NS and NS+R=R. One can also split and join arbitrary 1|1-cycles by first deforming them so that some parts of them become superconformal and then joining or splitting along these parts.

# 4 Semi-infinite forms and the superstring BRST complex

In this section we show that the superstring BRST complex can be constructed using the formalism of differential forms on supermanifolds. More specifically, the superstring BRST complex is the complex of semi-infinite forms on the super-Witt algebra with the values in a $\hat{c} = 10$ representation called the matter representation. The construction of semi-infinite forms on an infinite dimensional super Lie algebra is completely parallel to the corresponding construction in the bosonic case [38]. On the other hand the new ingredients brought in the odd directions to the de Rham complex are very well known for the superstring BRST complex [39, 40, 37, 26]. Therefore in the following we will simply present known results in a new light rather then obtain any new ones.

## 4.1 Semi-infinite forms

A super-Witt algebra, $\mathfrak{W}$, is simply the algebra of superconformal vector fields on a dimension 1|1 supercircle. Since there are two kinds of a supercircle, namely $S^{1|1}$ and $S^1 \times \mathbb{R}^{0|1}$, there are two types of super-Witt algebras. The first is known as Neveu-Schwarz and the second as Ramond algebra. If $\boldsymbol{t} = (t, \tau)$ parameterizes the supercircle in a way that $(t, \tau) = (t + 2\pi, \pm\tau)$ (the minus sign for the Neveu-Schwarz and plus for Ramond), we can write the generators of the super-Witt algebra explicitly as $-\frac{1}{2}$-differentials:

$$L_n = i\,e^{i\,n\,t}\,(\mathrm{d}\boldsymbol{t})^{-\frac{1}{2}} \tag{4.1.1}$$

$$G_r = 2i\,\tau\,e^{i\,n\,t}\,(\mathrm{d}\boldsymbol{t})^{-\frac{1}{2}}, \tag{4.1.2}$$

where $r$ is integer for the Ramond and half-integer for the Neveu-Schwarz case. Normalization of the generators is is chosen so that the commutators of the corresponding vector fields have the standard form

$$[L_m, L_n] = (m - n)\,L_{m+n}, \tag{4.1.3}$$

$$[L_m, G_r] = \left(\frac{1}{2}m - r\right)G_{m+r}, \tag{4.1.4}$$

$$[G_r, G_s] = 2\,L_{r+s}. \tag{4.1.5}$$

Let $L_n^\star$ and $G_n^\star$ be linear functions on the super-Witt algebra such that the only nonzero values on the generators (4.1.1) and (4.1.2) are

$$L_n^\star(L_{-n}) = G_r^\star(G_{-r}) = 1. \tag{4.1.6}$$

We can use $L'_n = \Pi L^\star_n$ and $G'_n = \Pi G^\star_n$ as an analog of $dx^i$ and $d\xi^\mu$ to construct semi-infinite forms. Semi-infinite forms are those that can be generated from a Plücker form given by

$$|\,p,q\,\rangle_{\text{gh}} = \prod_{n \geq p} L'_n \prod_{r \geq q} \delta(G'_r) \qquad (4.1.7)$$

applying a finite number of *ghost operators*

$$c_n = \mathbf{e}(L^\star_n) \qquad\qquad \gamma_r = \mathbf{e}(G^\star_r) \qquad (4.1.8)$$
$$b_n = \mathbf{i}(L_n) \qquad\qquad \beta_r = \mathbf{i}(G_r), \qquad (4.1.9)$$

and delta functions of them. Standard commutation relations,

$$[b_n, c_m]_+ = \delta_{n,-m} \qquad [\beta_r, \gamma_s]_- = \delta_{r,-s}, \qquad (4.1.10)$$

trivially follow from the definitions. The Plücker forms of the kind given by eq. (4.1.7) are called ghost vacua and satisfy the following conditions

$$b_n|\,p,q\,\rangle_{\text{gh}} = 0 \quad \text{for } n \geq p \qquad c_n|\,p,q\,\rangle_{\text{gh}} = 0 \quad \text{for } n > -p$$
$$\beta_r|\,p,q\,\rangle_{\text{gh}} = 0 \quad \text{for } r \geq q \qquad \gamma_r|\,p,q\,\rangle_{\text{gh}} = 0 \quad \text{for } r \geq -q$$

The space of semi-infinite forms, $\mathbf{\Omega}(\mathfrak{W})$, can generated from the ghost vacua applying a finite number of $b$, $c$, $\beta$ and $\gamma$ operators. Loosely speaking, $\mathbf{\Omega}(\mathfrak{W})$ is the union of de Rham complexes based on Plücker forms $|\,p,q\,\rangle_{\text{gh}}$.

It is common in superstring theory to describe ghost states in terms of their bosonisation. The correspondence between our notation for the ghost vacua and the notation of [37] is given by

$$|\,p,q\,\rangle_{\text{gh}} \leftrightarrow \;:\!e^{(2-p)\sigma} e^{(q-\frac{3}{2})\phi}\!:\,. \qquad (4.1.11)$$

## 4.2 The Lie derivative of the semi-infinite forms and the central extension

It is well known that a formal application of the definition of a Lie derivative to the case of semi-infinite forms leads to a divergence in $\mathcal{L}(L_0)$. One can overcome this by redefining $\mathcal{L}(L_0)$ through "subtraction of infinity", or "normal ordering".

Recall that the Lie derivative can be deduced from the commutation relations (4.1.3) and (4.1.5) and is given by

$$\mathcal{L}(L_n) = \mathcal{L}_{bc}(L_n) + \mathcal{L}_{\beta\gamma}(L_n) \qquad (4.2.1)$$
$$= \sum_{m=-\infty}^{\infty} (m - 2n)\, c_{n-m} b_m - \sum_{r=-\infty}^{\infty} \left(s - \frac{3}{2}\right) \gamma_{r-s} \beta_r, \qquad (4.2.2)$$

$$\mathcal{L}(G_r) = \sum_{m=-\infty}^{\infty} \gamma_{r-m} b_m + (3r - m) c_{r-m} \beta_m \qquad (4.2.3)$$

Since normal ordering modifies only $L_0$, we consider the bosonic generators first. Following [26], we can treat ghost ($\mathcal{L}_{bc}(L_n)$) and anti-ghost ($\mathcal{L}_{\beta\gamma}(L_n)$) parts of eq. (4.2.1) simultaneously. Let $\boldsymbol{b}$ and $\boldsymbol{c}$ stand either for $b$ and $c$ or for $\beta$ and $\gamma$ then

$$\mathcal{L}_{\boldsymbol{bc}}(L_n) = \sum_{m=-\infty}^{\infty} (m - J_{\boldsymbol{c}} n)\boldsymbol{b}_{n-m}\boldsymbol{c}_m = \sum_{m=-\infty}^{\infty} (m - J_{\boldsymbol{b}} n)\boldsymbol{c}_{n-m}\boldsymbol{b}_m, \quad (4.2.4)$$

where $J$ denotes the *spin* of the corresponding operator, $J_{\boldsymbol{b}} + J_{\boldsymbol{c}} = 1$,

$$\begin{array}{rclcrcl} J_b &=& 2, & & J_c &=& -1, \\ J_\beta &=& 3/2, & & J_\gamma &=& -1/2; \end{array} \quad (4.2.5)$$

and $\epsilon = 1$ for $bc$ and $\epsilon = -1$ for $\beta\gamma$. Ghost vacua can also be split into the ghost and anti-ghost part

$$|p,q\rangle_{\text{gh}} = |p\rangle_{bc} \otimes |q\rangle_{\beta\gamma}, \quad |p\rangle_{bc} = L'_p L'_{p+1} \cdots \quad |q\rangle_{\beta\gamma} = \delta(G'_q)\delta(G'_{q+1})\cdots \quad (4.2.6)$$

with $|q\rangle_{\boldsymbol{bc}}$ denoting either of them. The normal ordering is defined by

$$:\boldsymbol{c}_n \boldsymbol{b}_{-n}: = \begin{cases} \boldsymbol{c}_n \boldsymbol{b}_{-n} & \text{for } n < J_{\boldsymbol{c}} \\ -\epsilon\, \boldsymbol{b}_{-n} \boldsymbol{c}_n & \text{for } n > J_{\boldsymbol{c}}. \end{cases} \quad (4.2.7)$$

Normal ordered Lie derivatives considered as operators on the space of semi-infinite forms $\boldsymbol{\Omega}(\mathfrak{W})$ do not form a representation of $\mathfrak{W}$. Instead they form a representation $R : \mathfrak{Vir} \to \text{Aut}(V)$ of its non-trivial central extension—the *super Virasoro algebra* or $\mathfrak{Vir} = \mathfrak{vect}(S^{1|1}) \oplus \hat{c}\,\mathbb{C}$ with the commutation relations given by

$$[L_m, L_n] = (m - n)\, L_{m+n} + \frac{1}{8}\hat{c}(m^3 - m)\delta_{m,-n}, \quad (4.2.8)$$

$$[L_m, G_r] = \left(\frac{1}{2}m - r\right) G_{m+r}, \quad (4.2.9)$$

$$[G_r, G_s] = 2\, L_{r+s} + \frac{1}{2}\hat{c}\left(r^2 - \frac{1}{4}\right)\delta_{r,-s}. \quad (4.2.10)$$

The value of the *central charge* is determined by the difference of the ghost spins $Q_{\boldsymbol{bc}} = \epsilon(J_{\boldsymbol{b}} - J_{\boldsymbol{c}})$ which is also called the *background charge*

$$R_{\boldsymbol{bc}}(\hat{c}) = \frac{2}{3}\epsilon(1 - 3\, Q_{\boldsymbol{bc}}^2). \quad (4.2.11)$$

Substituting the values of the background charge $Q_{bc} = -3$ and $Q_{\beta\gamma} = 2$, we obtain the well known values

$$R_{bc}(\hat{c}) = \frac{2}{3}(-26), \qquad R_{\beta\gamma}(\hat{c}) = \frac{2}{3}(11), \qquad R(\hat{c}) = -10. \quad (4.2.12)$$

The ghost vacua are eigen states of $:\mathcal{L}_{bc}(L_0):$

$$:\mathcal{L}_{bc}(L_0):|\,q\,\rangle_{bc} = \frac{1}{2}\epsilon(q+J_b)(q+J_c)|\,q\,\rangle_{bc}; \qquad (4.2.13)$$

and its eigen values are also known as conformal weights. There are two forms with zero conformal weight, namely $|-J_b\rangle_{bc}$ and $|-J_c\rangle_{bc}$. The former is also annihilated by $L_{\pm 1}$ and therefore defines an $SL(2,\mathbb{C})$ vacuum.

## 4.3 Ghost number operator

The ghost number operator is also defined using the normal ordering

$$j_0 = \sum_{n=-\infty}^{\infty} :c_{-n}b_n: + \sum_{r=-\infty}^{\infty} :\gamma_{-r}\beta_r:. \qquad (4.3.1)$$

In the bosonic case the ghost number operator provides grading on the BRST complex which is a regularized version of the degree for the semi-infinite forms. In the super case forms have two degrees, even and odd. Both of them can be defined by fixing the degree of the $SL(2,\mathbb{C})$ vacuum, $|-J_b\rangle_{bc}$ as zero. The ghost number operator provides only the difference of the even and odd degree.

$$j_0 = \deg_{\bar{0}} - \deg_{\bar{1}}. \qquad (4.3.2)$$

The operators that calculate the degrees separately can be defined in terms of bosonized currents [37],

$$j_\sigma = \oint \partial\sigma, \quad j_\phi = \oint \partial\phi, \quad j_\chi = \oint \partial\chi, \qquad (4.3.3)$$

as follows

$$\deg_{\bar{0}} = j_\sigma - j_\chi \qquad (4.3.4)$$
$$\deg_{\bar{1}} = -j_\phi - j_\chi. \qquad (4.3.5)$$

Ghost vacua are uniquely specified by their even and odd degrees. This observation suggests an alternative notation for the ghost vacua which labels them by their dimension:

$$|\,r|s\,\rangle_{\text{gh}} = |\,r+2, s+\frac{3}{2}\,\rangle_{\text{gh}}. \qquad (4.3.6)$$

## 4.4 BRST operator

When we consider the action of de Rham differential on semi-infinite forms we encounter the same problem as for the Lie derivatives. In fact $\mathbf{d}$ can be easily expressed in terms of $\mathcal{L}(L_n)$ and $\mathcal{L}(G_r)$

$$\mathbf{d} = \frac{1}{2}\left(\sum_{n=-\infty}^{\infty} c_{-n}\mathcal{L}(L_n) + \sum_{r=-\infty}^{\infty} \gamma_{-r}\mathcal{L}(G_r)\right). \qquad (4.4.1)$$

The normal ordering which we used to define the Lie derivatives of semi-infinite forms brakes an important identity $\mathbf{d}^2 = 0$ since

$$:\mathbf{d}:^2 = \sum_{m,n=-\infty}^{\infty} ([:\mathcal{L}(L_m):,:\mathcal{L}(L_n):] - :\mathcal{L}([L_m, L_n]):)c_n c_m$$
$$+ \sum_{r,s=-\infty}^{\infty} ([:\mathcal{L}(G_r):,:\mathcal{L}(G_s):] - :\mathcal{L}([G_r, G_s]):)\gamma_s \gamma_r$$
$$= \sum_{n=-\infty}^{\infty} \frac{1}{8}\hat{c}(n^3 - n)\, c_n c_{-n} + \sum_{r=-\infty}^{\infty} \frac{1}{2}\hat{c}\left(r^2 - \frac{1}{4}\right)\gamma_r \gamma_{-r} \neq 0 \quad (4.4.2)$$

In order to have a nilpotent BRST operator, $Q$ one has to consider semi-infinite forms with the values in a representation $\mathcal{R}$ of the super-Virasoro algebra with $\hat{c} = 10$.

## 4.5 GSO projection and the chiral BRST complex

The supercircle used in our approach doesn't have the orientation in the odd direction. Moreover, in the Neveu-Schwarz sector this orientation cannot be defined even in principle. Therefore we have to restrict the complex of semi-infinite forms to those that are invariant under the flip of the odd direction, $\tau \mapsto -\tau$, or equivalently $G_r \mapsto -G_r$ this procedure is known as the GSO projection. Note that since a semi-infinite form contains a product of infinite number of $\delta(G_r)$, the result of the flip is ill defined. This can be fixed by choosing an invariant vacuum since then the sign of any other state can be deduced by a finite procedure. In the Neveu-Schwarz sector a natural choice for the invariant vacuum is the $SL(2, \mathbb{C})$ vacuum, but in the Ramond sector there are two possibilities. This becomes important when two chiral complexes are combined to define a complex for the type II superstring. Depending on what choice was made one would obtain a type IIA or type IIB theory as a result. Finally we can define the chiral BRST complex as a GSO projection of $\mathbf{\Omega}(\mathfrak{W}, \mathcal{R})$

$$\mathcal{E} = [\mathbf{\Omega}(\mathfrak{W}, \mathcal{R})]_{\text{GSO}}. \quad (4.5.1)$$

# 5 Superstring backgrounds

In this section we will define a notion of string background based on that of a topological conformal field theory [41]. Our approach to TCFT will be based on the differential forms on the moduli space of decorated superconformal surfaces, $\mathcal{P}$ and BRST cohomology. The transition from TCFTs to string backgrounds will be made by replacing $\mathcal{P}$ with $\hat{\mathcal{P}}$ (the space of surfaces similar to $\mathcal{P}$, but without marked points on the boundaries) and replacing the BRST cohomology with its semi-relative analog.

## 5.1 Topological conformal field theory. Purely even case.

Let us start by recalling the definition of topological conformal field theory in the purely even or bosonic case. We define a decorated conformal surface as a two-dimensional surface with a boundary equipped with a complex structure and a parameterization of each connected component of the boundary. We will now call the connected components of the boundary just boundaries for short. Thus a boundary will mean one of the boundaries. If the orientation of a boundary given by the parameterization agrees with the natural orientation of the surface, it is called in-boundary, otherwise it is called out-boundary.

The space of decorated conformal surfaces with $n$ in and $m$ out-boundaries will be denoted by $\mathcal{P}_{n,m}$. Sewing operation $\bowtie$ takes two surfaces with the number of out-boundaries of the first being the same as the number of in-boundaries of the second and yields a new surface by identifying the points of these boundaries according to their parameterization.

$$\bowtie : \mathcal{P}_{n,m} \times \mathcal{P}_{m,l} \to \mathcal{P}_{n,l}. \tag{5.1.1}$$

To define a topological conformal field theory or TCFT one has to provide the following data [41]

1. a differential complex $(\mathcal{E}, Q)$, where $\mathcal{E}$ is a graded vector space and $\boldsymbol{Q}$ a nilpotent operator of degree 1;

2. a differential form $\omega(m,n)$ on $\mathcal{P}_{n,m}$ with the values in $\mathrm{Hom}(\mathcal{E}^{\otimes m}, \mathcal{E}^{\otimes m})$ for each $m$ and $n$.

These data have to satisfy certain conditions, the most important for us will be that $\omega(m,n)$ is closed of the total degree $q(n-\chi/2)$, where $q$ is the anomaly charge and $\chi$ the Euler number of the surface:

$$(\mathrm{d} + Q)\omega = 0 \tag{5.1.2}$$

and the sewing operation on $\mathcal{P}$ corresponds to the external multiplication of the operator-valued forms.

In our discussion differential complex will always be the BRST complex, which in the bosonic case has anomaly $q = -3$. The differential will be the BRST operator and the grading will be provided by the ghost number. The differential forms will be reconstructed from a combined conformal field theory of ghosts and matter using the standard procedure. We will briefly review this procedure when we return to the superstring.

Since $\omega(m,n)$ is closed, its integral over a cycle $C$ defines an automorphism in cohomology of $(\mathcal{E}, Q)$ which only depends on the homology class of $C$

$$\boldsymbol{A}_C = \int_C \omega(m,n) : H(\mathcal{E}, Q)^{\otimes m} \to H(\mathcal{E}, Q)^{\otimes n}. \tag{5.1.3}$$

The degree of $\boldsymbol{A}_C$ is

$$|\boldsymbol{A}_C| = 3(\chi/2 - n) - \dim C. \tag{5.1.4}$$

Thus a zero-cycle $C_\bullet$ in $\mathcal{P}_{2,1}$ consisting of a sphere with three holes, defines a multiplication in $H(\mathcal{E}, Q)$

$$\phi \cdot \psi = \boldsymbol{A}_{C_\bullet}(\phi \otimes \psi). \tag{5.1.5}$$

This cycle defines a unique homology class and thus the multiplication is well defined. The degree formula (eq. (5.1.4)) implies that

$$|\phi \cdot \psi| = |\phi| + |\psi|. \tag{5.1.6}$$

Another important operator is provided by a one-cycle in $\mathcal{P}_{1,1}$ consisting of spheres with two holes which differ only by the position of the starting point on the out-boundary. We will call this cycle $C_\Delta$, where $\Delta$ is the corresponding operator of degree $-1$. It is possible to fix $C_\Delta$ completely (not up-to homology) by choosing it to run over degenerate surfaces made of two coincident boundaries. This choice will define operator $\Delta$ unambiguously on the complex $\mathcal{E}$ rather than on its cohomology. From the sewing properties of $C_\Delta$ one can deduce that $\Delta^2 = 0$.

Finally, we define a one-cycle $C_{\text{BV}}$ in $\mathcal{P}_{2,1}$ by rotating one of the holes around the equator while keeping the other two fixed at the poles. The correspondent operator defines the BV bracket

$$\{\phi, \psi\} = \boldsymbol{A}_{C_{\text{BV}}}(\phi \otimes \psi) \tag{5.1.7}$$

of degree $-1$.

One dimensional homology of $\mathcal{P}_{2,1}$ can be generated by three cycles $C_i$ where the first two are obtained by rotating one of the in-boundaries and the last by rotating the out boundary. Accordingly, $C_{\text{BV}}$ can be represented as

$$C_{\text{BV}} = C_3 - C_2 + C_1 + \partial X, \tag{5.1.8}$$

which leads to the identity

$$\{\phi, \psi\} = \Delta(\phi \cdot \psi) - \Delta\phi \cdot \psi + \phi \cdot \Delta\psi \tag{5.1.9}$$

in BRST cohomology.

## 5.2 Topological superconformal field theory.

At this point we have all the tools available to generalize the notion of topological conformal field theory to the super case. Due to the lack of a better name we will call the resulting theory topological superconformal field theory and abbreviate it as TSFT. The generalization is straightforward.

The space of decorated conformal surfaces becomes the space of decorated *superconformal* surfaces, where decoration would mean a superconformal parameterization of the boundaries. In addition to their orientation, the boundaries can now be distinguished by their type as Neveu-Schwarz or Ramond.

The complex $(\mathcal{E}, Q)$ becomes a superspace with two gradings, $\deg_{\bar{0}}$ and $\deg_{\bar{1}}$ and $Q$ increases $\deg_{\bar{0}}$ by one while keeping $\deg_{\bar{1}}$ the same. Note that $\deg_{\bar{0}, \bar{1}}$ gradings are defined independently from the $\mathbb{Z}_2$-grading of $\mathcal{E}$. The $\mathbb{Z}_2$-grading splits $\mathcal{E}$ into $\mathcal{E}_{\bar{0}}$ and $\mathcal{E}_{\bar{1}}$, or Neveu-Schwarz and Ramond sectors, each having two gradings $\deg_{\bar{0}}$ and $\deg_{\bar{1}}$. It will be convenient to assume that $\deg_{\bar{1}}$ has half-integer values on $\mathcal{E}_{\bar{1}}$.

The differential forms $\omega(m,n)$ are now forms on the supermanifolds $\mathcal{P}_{m,n}$ with the values in double-graded space $\mathrm{Hom}(\mathcal{E}^{\otimes m}, \mathcal{E}^{\otimes m})$. Accordingly, these forms will have even and odd degree, or $\deg_{\bar{0}}$ and $\deg_{\bar{1}}$. We will use the symbol $\deg = \deg_{\bar{0}} | \deg_{\bar{1}}$ to simplify the notation. In this work we will only consider TSFTs based on the BRST complex with $-3|-2$-anomaly and thus the degree of $\omega(m,n)$ will be given by

$$\deg \omega(m,n) = 3|2\,(\chi/2 - n), \tag{5.2.1}$$

where $\chi$ is again the Euler number. The closeness condition, eq. (5.1.2) will remain the same

$$(\mathrm{d} + Q)\,\omega = 0, \tag{5.2.2}$$

but the differentials d and $Q$ will now have degree $1|0$.

The simplest way to obtain a TSFT is to add superconformal ghosts to a superconformal theory with the central charge $\hat{c} = 10$. Recall that a superconformal theory defines an operator valued *function* $\mathcal{F}(m,n)$ rather than differential form on $\mathcal{P}_{m,n}$. In order to obtain an $r|s$ form we need to add $r$ $B$-insertions and $s$ $\delta(B)$-insertions:

$$\omega(m,n)[\Sigma;\, v_1, \cdots, v_r, \nu_1, \cdots, \nu_s] =$$
$$\mathcal{F}(m,n)[\Sigma]\,B(v_1)\cdots B(v_r)\delta(B(\nu_1))\cdots\delta(B(\nu_r)), \tag{5.2.3}$$

where $\Sigma \in \mathcal{P}_{m,n}$ is a point in $\mathcal{P}_{m,n}$, and $v \in (T_\Sigma \mathcal{P}_{m,n})_{\bar{0}}$ and $\nu \in (T_\Sigma \mathcal{P}_{m,n})_{\bar{1}}$ are the even and odd tangent vectors at the point $\Sigma$. The construction of $B(v)$ is based on the Schiffer representation of the tangent space of $\mathcal{P}$ and is completely analogous to that in the pure even case. It will be useful to represent $\omega(m,n)$ in pseudoform-like notation as

$$\omega(m,n)[\Sigma, \mathrm{d}\Sigma] = \mathcal{F}(m,n)[\Sigma]\exp(B(\mathrm{d}\Sigma)). \tag{5.2.4}$$

Formally, eq. (5.2.3) is the Baranov-Schwarz transformation of eq. (5.2.4). The pseudoform-like representation is much easier for calculations than the Lagrangian representation. For example, it takes just a few lines to check that $\omega(m,n)$ is closed using eq. (5.2.4) (see [3]).

Different algebraic operations on the cohomology space $H(\mathcal{E}, Q)$ can be constructed in complete analogy with the purely even case using integrals over genus zero cycles $C$ in $\mathcal{P}_{2,1}$, yet due to the presence of the odd directions we have a reacher supply of cycles. For example one can use cycles of dimension $1|0$ and $1|2$ in order to define two different variants of $\Delta$ and $\{\cdot,\cdot\}$. The two variants of $\Delta$ can be related by the picture-changing operators. It will be convenient to define a universal operator $\Delta$ which acts differently in different sectors

$$|\Delta| = \begin{cases} 1|0 & \text{on } \mathcal{E}_{\bar{0}} \\ 1|1 & \text{on } \mathcal{E}_{\bar{1}} \end{cases} \qquad (5.2.5)$$

We define the $C_\bullet$ cycle in $\mathcal{P}_{2,1}$ to be Similarly we define the $C_{\mathrm{BV}}$ cycle to be 1|0-dimensional if the out-boundary is of NS-type, 1|1-dimensional if it is of R-type. Under these assumptions the identity given by eq. (5.1.9) will hold for the superstring BRST cohomology.

Note that since the contours have different dimensions, the bracket will have different degree.

$$|\{\phi,\psi\}| = \begin{cases} |\phi| + |\psi| - 1|0 & \text{if } \phi, \psi \in H(\mathcal{E}_{\bar{1}}, Q) \\ |\phi| + |\psi| - 1|1 & \text{otherwise.} \end{cases} \qquad (5.2.6)$$

## 5.3 BRST complex of the type II superstring

The BRST complex of the type II superstring is simply the tensor product of two copies of the chiral BRST complex $(\mathcal{E}, Q)$,

$$\boldsymbol{\mathcal{E}} = \mathcal{E} \otimes \mathcal{E}, \quad \boldsymbol{Q} = Q \otimes 1 + 1 \otimes Q. \qquad (5.3.1)$$

Operators acting only on the first factor are called chiral and those acting on the second factor antichiral. For such operators we will often use a short notation

$$O = O \otimes 1 \quad \text{and} \quad \overline{O} = 1 \otimes O. \qquad (5.3.2)$$

For example $\boldsymbol{Q}$ can be written as

$$\boldsymbol{Q} = Q + \overline{Q}. \qquad (5.3.3)$$

Since $\mathcal{E}$ has two sectors, Neveu-Schwarz and Ramond, the full type II BRST complex $\boldsymbol{\mathcal{E}}$ has four: NN, NR, RN and RR. NN and RR sectors form the even and NR and RN the odd subspace of $\boldsymbol{\mathcal{E}}$.

$$\boldsymbol{\mathcal{E}} = \boldsymbol{\mathcal{E}}_{\bar{0}} \oplus \boldsymbol{\mathcal{E}}_{\bar{1}}, \qquad (5.3.4)$$
$$\boldsymbol{\mathcal{E}}_{\bar{0}} = \mathcal{E}_{\bar{0}} \otimes \mathcal{E}_{\bar{0}} \oplus \mathcal{E}_{\bar{1}} \otimes \mathcal{E}_{\bar{1}}, \qquad (5.3.5)$$
$$\boldsymbol{\mathcal{E}}_{\bar{1}} = \mathcal{E}_{\bar{1}} \otimes \mathcal{E}_{\bar{0}} \oplus \mathcal{E}_{\bar{0}} \otimes \mathcal{E}_{\bar{1}}. \qquad (5.3.6)$$

## 5.4 Superstring backgrounds

Recall that the closed string backgrounds were defined as a set differential forms $\hat{\omega}(n)$ on $\hat{\mathcal{P}}_n$ with the values in $\mathrm{Hom}(\hat{\boldsymbol{\mathcal{E}}}^{\otimes n}, \mathbb{C})$ [42]. Here the space $\hat{\mathcal{P}}_n$ is the quotient space of $\mathcal{P}_{n,0}$ with respect to the action of $(S^1)^n$ which shifts the initial point on each of $n$ boundaries. Taking the quotient is equivalent to forgetting about the position initial points on the boundaries. Furthermore, the quotient space has to be completed to a bundle over $\overline{\mathcal{M}}_n$, the compactified moduli space of $n$-punctured surfaces. (The projection is realized by gluing the unit disks to each boundary). The space $\hat{\mathcal{E}}$ is the semi-relative BRST complex, or the subcomplex of $\mathcal{E}$ on which $L_0^- = b_0^- = 0$. Clearly any TCFT based on a BRST complex would define a closed string background. The perturbative spectrum of the physical states near the background is given by the ghost number two semi-relative BRST cohomology, $H^2(\hat{\mathcal{E}}, \boldsymbol{Q})$.

The only non-trivial step in generalizing this construction to superstrings is to realize that the initial point on a Ramond boundary has both even and odd coordinate. Therefore the super generalization of $\hat{\mathcal{P}}_n$ is the quotient $\mathcal{P}_{n,0}$ with respect to the action of $n_{\text{NS}}$ copies of $S^1$ and $n_{\text{R}}$ copies of $S^1 \times \mathbb{R}^{0|1}$, where $n_{\text{NS}}$ are $n_{\text{R}}$ respectively the numbers of Neveu-Schwarz and Ramond boundaries. Therefore the Ramond-Ramond sector of $\hat{\mathcal{E}}$ consists of the vectors from $\mathcal{E}_{\text{RR}}$ annihilated by $G_0^-$ and $\beta_0^-$. Note that $L_0^- = b_0^- = 0$ is automatically satisfied since $L_0^- = (G_0^-)^2$ and $b_0^- = [G_0^-, \beta_0^-]$. The perturbative spectrum of the physical states near the superstring background is given by the ghost number $2|1$ and $2|2$ cohomology of $\hat{\mathcal{E}}_{\bar{0}}$ (NS-NS and R-R sectors), and ghost number $2|1\frac{1}{2}$ cohomology of $\hat{\mathcal{E}}_{\bar{1}}$ (NS-R sector).

Additional condition, $\beta_0^- = 0$ in the Ramond sector results in a new family of Ramond-Ramond physical states which have the form

$$(c_1 \gamma_{-1} - \bar{c}_1 \bar{\gamma}_{-1}) |\psi\rangle = \boldsymbol{Q} \gamma_0^- |\psi\rangle, \qquad (5.4.1)$$

where $\boldsymbol{Q}|\psi\rangle = 0$ and $\deg |\psi\rangle = 0|2$. Although these states have the form $\boldsymbol{Q}|\lambda\rangle$, the are not necessarily trivial since $|\lambda\rangle = \gamma_0^- |\psi\rangle$ does not belong to the semi-relative complex. The significance of these additional Ramond-Ramond physical states is yet to be understood.

A standard procedure to construct a superstring background is to define a TSFT by combining a $\hat{c} = 10$ superconformal field theory with superconformal ghosts and then use two copies of this TSFT to construct a background. These two copies can be chosen differently to provide a large spectrum of superstring backgrounds.

## 5.5 Superstring amplitudes

Our construction of the superstring amplitudes will be based on the following statement. There exists a global non-holomorphic section $\sigma$ of $\hat{\mathcal{P}}_n$ over $\overline{\mathcal{M}}_n$ and all such sections are homologically equivalent. An explicit construction of such section was done in [15]. The second provision of the statement is based on the fact $\hat{\mathcal{P}}_n$ is homotopy equivalent to $\overline{\mathcal{M}}_n$ and therefore all the sections are homotopic to each other and thus homologically equivalent.

Integrating $\hat{\omega}(n)$ over $\sigma(\overline{\mathcal{M}}_n)$, we obtain a map

$$\boldsymbol{A}_{\sigma(\overline{\mathcal{M}}_n)} = \int_{\sigma(\overline{\mathcal{M}}_n)} \hat{\omega}(n) : \boldsymbol{\mathcal{E}}^{\otimes n} \to \mathbb{C}. \qquad (5.5.1)$$

known as the superstring amplitude. Since $\hat{\omega}(n)$ is closed, the string amplitude on physical states does not depend neither on the choice of $\sigma$ nor on the choice of the representatives.

$$\boldsymbol{A}_{\sigma(\overline{\mathcal{M}}_n)}(\Psi) = \boldsymbol{A}_{\sigma'(\overline{\mathcal{M}}_n)}(\Psi), \tag{5.5.2}$$

where $\sigma$ and $\sigma'$ are two different sections; and

$$\boldsymbol{A}_{\sigma(\overline{\mathcal{M}}_n)}(\Psi + \boldsymbol{Q}\Lambda) = \boldsymbol{A}_{\sigma(\overline{\mathcal{M}}_n)}(\Psi + \boldsymbol{Q}\Lambda), \tag{5.5.3}$$

where $\Psi, \Lambda \in \hat{\boldsymbol{\mathcal{E}}}^{\otimes n}$. In other words it defines a unique $n$-linear form $A$ on $H(\hat{\boldsymbol{\mathcal{E}}}, \boldsymbol{Q})$

$$A: H(\hat{\boldsymbol{\mathcal{E}}}, \boldsymbol{Q})^{\otimes n} \to \mathbb{C}. \tag{5.5.4}$$

# 6 Applications

In this final section we will present a few applications of the new formalism. One of the great advantages that we gain from our approach is that we can use arbitrary representatives in the BRST classes of physical states in order to calculate the amplitude. This allows us to use representatives which are not given by primary states. This is particularly important in the case when there is now primary representative in the BRST class. This is the case with the dilaton. We will present a straightforward calculation which leads to the dilaton theorem (or rather its ghost part). The second example will be the relation between the symmetries of the background and and the BRST cohomology at a particular ghost number. The symmetries generated by this cohomology will be exact in all orders in perturbation theory. Interestingly, those are quantum or stringy symmetries which may or may not coincide with the classical symmetries of the background. Finally we will present the symmetry generators for the case of the flat $D = 10$ background and show that they form the $D = 10$ super Poincaré algebra.

## 6.1 Dilaton theorem

Dilaton state at momentum $p$ ($p^2 = 0$) can be represented by the following closed state state in the BRST complex

$$|\,D(p)\,\rangle' = \eta^\perp_{\mu\nu} \psi^\mu_{-1/2} \bar{\psi}^\nu_{-1/2} e^{i\,p\cdot x} |\,2|2\,\rangle_{\mathbf{gh}}. \tag{6.1.1}$$

This state is a combination of the standard dilaton vertex [43] with the combined chiral-antichiral ghost vacuum of total degree $2|2$ (*c.f.* [44]), where the later can be related to the $\mathrm{SL}(2,\mathbb{C})$ ghost vacuum as follows

$$|\,2|2\,\rangle_{\mathbf{gh}} = |\,1|1\,\rangle_{\mathrm{gh}} \otimes |\,1|1\,\rangle_{\overline{\mathrm{gh}}} = c_1 \bar{c}_1 \delta(\gamma_{1/2}) \delta(\bar{\gamma}_{1/2}) |\,0\,\rangle. \tag{6.1.2}$$

The symbol $\eta^\perp_{\mu\nu}$ denotes the projector onto the space transverse to $p$ and can be written explicitly using an auxiliary vector $\tilde{p}$ such that $\tilde{p}^2 = 0$ and $p \cdot \tilde{p} = 1$

$$\eta^\perp_{\mu\nu} = \eta_{\mu\nu} - p_\mu \tilde{p}_\nu - p_\nu \tilde{p}_\mu. \tag{6.1.3}$$

One can easily show that the cohomology class of $|D(p)\rangle$ does not depend on the choice of $\tilde{p}$. Indeed, if we add the BRST exact state,

$$\boldsymbol{Q}(\beta_{1/2}\tilde{p} \cdot \psi_{-1/2} - \bar{\beta}_{1/2}\tilde{p} \cdot \bar{\psi}_{-1/2}) e^{i p \cdot x} |2|2\rangle_{\mathbf{gh}} \tag{6.1.4}$$

we can find another representative, $|D(p)\rangle$, of the same cohomology class which has no $\tilde{p}$ in it

$$|D(p)\rangle = \left[\psi_{-1/2} \cdot \bar{\psi}_{-1/2} - (\gamma_{-1/2}\bar{\beta}_{-1/2} - \bar{\gamma}_{-1/2}\beta_{-1/2})\right] e^{i p \cdot x} |2|2\rangle_{\mathbf{gh}}. \tag{6.1.5}$$

This new representative for the dilaton is more useful because it is manifestly Lorenz covariant. An extra ghost piece in eq. (6.1.5) is called *ghost dilaton*. Ghost dilaton is annihilated by the BRST operator only at $p = 0$ and defines a discrete physical state. This state,

$$|D_{\mathrm{gh}}\rangle = (\gamma_{-1/2}\bar{\beta}_{-1/2} - \bar{\gamma}_{-1/2}\beta_{-1/2}) |2|2\rangle_{\mathbf{gh}}, \tag{6.1.6}$$

will be the subject of the following analysis.

The ghost dilaton may appear to be trivial since

$$|D_{\mathrm{gh}}\rangle = \boldsymbol{Q}(c_0 - \bar{c}_0)\beta_{-\frac{1}{2}}\bar{\beta}_{-\frac{1}{2}} |2|2\rangle_{\mathbf{gh}} = \boldsymbol{Q}|\chi\rangle, \tag{6.1.7}$$

but it is not because $|\chi\rangle$ is not annihilated by $b_0^-$ and therefore does not belong to the semi-relative BRST complex. The identity (6.1.7) will play a crucial role in our approach to the dilaton theorem.

The (ghost) dilaton theorem reduces an amplitude of one ghost dilaton with other physical states to that without the dilaton. Let us denote the $g$-loop amplitude in the presence of the ghost dilaton by $A_D(\Psi)$. Our goal is to express $A_D(\Psi)$ in terms of $A(\Psi)$. According to eqs. (5.5.1) and (5.2.4),

$$A_D(\Psi) = \int_{\sigma \overline{\mathcal{M}}_{g,n+1}} \langle \Sigma_{g,n+1}(x) | e^{B(\mathrm{d}\Sigma)} | \Psi \rangle \otimes | D_{\mathrm{gh}} \rangle, \tag{6.1.8}$$

where $\langle \Sigma_{g,n+1} | = \mathcal{F}(n+1,0)[\Sigma_{g,n+1}]$ (see also [3]). The section $\sigma : \overline{\mathcal{M}}_{g,n+1} \to \hat{\mathcal{P}}_{g,n+1}$ or equivalently the local coordinates around the punctures, can be selected arbitrarily, but it will be convenient to make the local coordinates around the $n$ punctures where the states from $\Psi$ are inserted independent on the position of the puncture with the dilaton. This choice would allow us to factor out the integral over the position of the dilaton insertion from the amplitude. Unfortunately when the dilaton insertion comes close to one of the other punctures, we have to begin changing both of their local coordinates in order to avoid the overlap. This situation can be better described by placing the dilaton and the other state on a three punctured sphere $S_3$ with a fixed choice of local coordinates and pitching this sphere out of the surface. Let $\mathcal{F}_{g,n+1}$ be the vicinity of the compactification divisor in $\overline{\mathcal{M}}_{g,n+1}$ which consists of surfaces which can be obtained this way and let $\mathcal{V}_{g,n+1} = \overline{\mathcal{M}}_{g,n+1} \backslash \mathcal{F}_{g,n+1}$. Note that since the positions of the punctures on $S_3$ are fixed, we can fix the phase of the local coordinate near the dilaton insertion over $\mathcal{F}_{g,n+1}$. This will allow the insertion of $|\chi\rangle$ and make the integrand in eq. (6.1.8) an exact form over $\mathcal{F}_{g,n+1}$. Using the Stokes theorem, we can rewrite the amplitude $A_D(\Psi)$ as follows

$$A_D(\Psi) = \int_{\sigma(\mathcal{V}_{g,n+1})} \langle \Sigma_{g,n+1}(x) | e^{B(\mathrm{d}\Sigma)} | \Psi \rangle \otimes | D_{\mathrm{gh}} \rangle$$
$$+ \int_{\sigma(\partial \mathcal{V}_{g,n+1})} \langle \Sigma_{g,n+1}(x) | e^{B(\mathrm{d}\Sigma)} | \Psi \rangle \otimes | \chi \rangle. \quad (6.1.9)$$

Now we can choose the section $\sigma$ so that the integral over the position of the dilaton insertion factors out and find that

$$A_D(\Psi) = \left( \int_{\hat{\Sigma}_{g,n}} \omega_D - \int_{\partial \hat{\Sigma}_{g,n}} \omega_\chi \right) A(\Psi). \quad (6.1.10)$$

where $\omega_\chi$ and $\omega_D$ are the differential forms corresponding to the insertion of $|\chi\rangle$ and $|D_{\mathrm{gh}}\rangle$ respectively,

$$B(v)\,\delta(B(\nu))\,|\chi\rangle = \omega_\chi(v|\nu)|0\rangle \quad (6.1.11)$$

and

$$B(v_1)\,B(v_2)\,\delta(B(\nu_1))\,\delta(B(\nu_2))\,|D_{\mathrm{gh}}\rangle = \omega_D(v_1,v_2|\nu_1,\nu_2)|0\rangle. \quad (6.1.12)$$

Explicit calculation of the factor in eq. (6.1.10) is done in appendix A and the final result is

$$A_D(\Psi) = (2 - 2g - n)\,A(\Psi). \quad (6.1.13)$$

Note that we did not make any assumption about the nature of the physical states which form $\Psi$. One or more of them can also be the ghost dilaton.

It worth to mention that the Euler characteristic which appears as a factor in the dilaton equation does not depend on the type of the states involved in the amplitude. This seems to be in contradiction to the fact that the dilaton couples differently to the Ramond and Neveu-Schwarz states [45]. A careful analysis shows that the unusual RR-dilaton coupling should be attributed to the matter part of the dilaton, $|D_{\mathrm{m}}\rangle = \psi_{-1/2} \cdot \bar{\psi}_{-1/2} |2|2\rangle_{\mathrm{gh}}$.

## 6.2 Cohomology and the symmetries of the background

In the bosonic string theory global symmetries of the background can be identified with the ghost number one semi-relative BRST cohomology at zero momentum. Similar statement is true for the superstring. A new ingredient which is specific for the superstring is the two type of symmetries, even and odd. Even symmetries map bosons to bosons and fermions to fermions while the odd symmetries interchange bosons and fermions. We are going to show that the even symmetries can be identified with $H^{1|2}(\hat{\mathcal{E}}_{\mathrm{NN}}, \boldsymbol{Q}) \oplus H^{1|1}(\hat{\mathcal{E}}_{\mathrm{RR}}, \boldsymbol{Q})$ while odd symmetries are given by $H^{1|\frac{3}{2}}(\hat{\mathcal{E}}_{\mathrm{NR}}, \boldsymbol{Q}) \oplus H^{1|\frac{3}{2}}(\hat{\mathcal{E}}_{\mathrm{RN}}, \boldsymbol{Q})$.

First recall how this works in the bosonic case. With every $\boldsymbol{Q}$-closed state $J$ of ghost number one we associate a transformation $\delta_J$ on the physical states, given by the formula

$$\delta_J \psi = \{J, \psi\}. \qquad (6.2.1)$$

Since the degree of the BV bracket is $-1$ this transformation does not change the ghost number of the state. Form the following series of identities

$$0 = \int_{\sigma(\overline{\mathcal{M}}_{n+1} \backslash D_\epsilon)} \hat{\omega}(n) \Psi \otimes \boldsymbol{Q} J \qquad \text{since } \boldsymbol{Q} J = 0 \qquad (6.2.2)$$

$$= \int_{\sigma(\overline{\mathcal{M}}_{n+1} \backslash D_\epsilon)} \mathrm{d}\hat{\omega}(n+1) \Psi \otimes J \qquad \text{by the closeness of } \hat{\omega}(n) \qquad (6.2.3)$$

$$= \int_{\sigma(\partial(\overline{\mathcal{M}}_{n+1} \backslash D_\epsilon))} \hat{\omega}(n+1) \Psi \otimes J \qquad \text{by the Stokes theorem} \qquad (6.2.4)$$

$$= \sum_{i=1}^{n} \int_{\sigma(\overline{\mathcal{M}}_n) \underset{i}{\bowtie} C_\bullet} \hat{\omega}(n+1) \Psi \otimes J \qquad \text{by contour deformation} \qquad (6.2.5)$$

$$= \sum_{i=1}^{n} \int_{\sigma(\overline{\mathcal{M}}_n)} \hat{\omega}(n) \delta_J^{(i)} \Psi \qquad \text{by the sewing axiom and eq. (5.1.9)} \qquad (6.2.6)$$

$$= \sum_{i=1}^{n} A(\delta_J^{(i)} \Psi) \qquad \text{by eq. (5.5.1)} \qquad (6.2.7)$$

we conclude that the amplitudes are invariant under $\delta_J$ transformation. The following notation was used: $D_\epsilon$ for the subspace of $\overline{\mathcal{M}}_{n+1}$ consisting of surfaces with the $n+1$-st puncture being infinitesimally close to one of the other $n$ punctures; $\Psi$ for an arbitrary element of $\hat{\mathcal{E}}^{\otimes n}$ and $\underset{i}{\bowtie}$ for the twist-sewing along the $i$-th boundary.

When all the ingredients in the eqs. (6.2.1–6.2.7) are properly generalized to the case of superstrings, these same equations can be used to construct the symmetry generators on for the superstring states. The dimensions for the contours used in the definition of the BV bracket are dictated by the dimension of the moduli space of superconformal surfaces with different types of punctures. Contour deformation argument should be carried out carefully for all different types of punctures taking into account that two Ramond punctures share common odd coordinate in the moduli space.

For $J \in \hat{\mathcal{E}}_{\bar{0}}$, the situation is completely analogous to the bosonic case. The degree of $J$ has to be $1|2$, the same as the dimension of the contour in the definition of the BV bracket. Therefore an even symmetry $\delta_J$ doesn't change the ghost number.

An odd symmetry has to transform bosonic states (with integer odd degree) to fermionic states (with half-integer odd degree). One can see that the correct degree for the odd symmetry generator is $1|\frac{3}{2}$.

## 6.3 Space-time SUSY without picture changing

In the previous section we have shown that the generator of symmetries of the superstring backgrounds can be obtained from the BRST cohomology. Even generators where identified with ghost number $1|1$ NS-NS cohomology states and odd generators with ghost number $1|\frac{1}{2}$, NS-R.

$$|P^\mu\rangle = \psi^\mu_{-\frac{1}{2}} |1|1\rangle \otimes |0|1\rangle \tag{6.3.1}$$

$$|Q_\alpha\rangle = \Sigma_\alpha |1|\frac{1}{2}\rangle \otimes |0|1\rangle \tag{6.3.2}$$

$$|\overline{Q}_\alpha\rangle = |0|1\rangle \otimes \Sigma_\alpha |1|\frac{1}{2}\rangle \tag{6.3.3}$$

The antichiral version of the state $|P^\mu\rangle$ is BRST equivalent to the chiral one [46]. Calculating the BV brackets of these states we obtain the commutation relations for the super-Poincaré algebra.

$$\{Q_\alpha, Q_\beta\} = \{\overline{Q}_\alpha, \overline{Q}_\beta\} = \Gamma^\mu_{\alpha\beta} P_\mu \tag{6.3.4}$$

## A  Integrating the dilaton and $\chi$ insertions

In this appendix we present the calculations showing that the sum of two integrals that appears in the ghost dilaton theorem is indeed equal to the Euler character of the surface.

$$\int_{\hat{\Sigma}_{g,n}} \omega_D - \int_{\partial \hat{\Sigma}_{g,n}} \omega_\chi = 2 - 2g - n. \tag{A.1}$$

First of all we have to calculate the differential forms $\omega_D$ and $\omega_\chi$, which were defined as the insertions of the dilaton and $\chi$ state respectively

$$B(v)\,\delta(B(\nu))\,|\chi\rangle = \omega_\chi(v|\nu)|0\rangle \tag{A.2}$$

and

$$B(v_1)\,B(v_2)\,\delta(B(\nu_1))\,\delta(B(\nu_2))\,|D_{\text{gh}}\rangle = \omega_D(v_1, v_2|\nu_1, \nu_2)|0\rangle. \tag{A.3}$$

These forms depend on the choice of local coordinates around the point of insertions, which can be described using an auxiliary coordinate $\boldsymbol{z}$ by a superconformal map $\boldsymbol{z} = H(\boldsymbol{w})$, where $H = (h, \hat{h})$ depends implicitly on the moduli parameters $t^A$, $A = 1, \ldots, 2|2$ associated with the position of the insertion point. Later we will specify these parameters to be given by the position of the insertion point in the $\boldsymbol{z}$ coordinate, i.e. $t^A = (h(0), \bar{h}(0); \hat{h}(0), \hat{\bar{h}}(0))$, but meanwhile it can be assumed to be arbitrary.

Let $v = \sigma_* \mathrm{d}t^A$ be the Schiffer variation associated with $\mathrm{d}t^A$, then in the uniformizing coordinate

$$v(\boldsymbol{z}) = \sigma_* \mathrm{d}t^A = \mathrm{d}_{t^A} h(\boldsymbol{w}) + \theta \, \mathrm{d}_{t^A} h(\boldsymbol{w}), \tag{A.4}$$

where

$$\mathrm{d}_{t^A} = \sum_{A=1}^{2|2} \mathrm{d}t^A \frac{\partial}{\partial t^A}. \tag{A.5}$$

On the other hand, in order to use it in eqs. (A.2) and (A.3), the Schiffer vector has to be transformed to the local coordinate $\boldsymbol{w}$. Since the Schiffer variation is a superconformal vector field, it transforms as a $(-1/2, 0)$ superconformal field (see eq. (3.1.4)), and

$$v(\boldsymbol{w}) = \frac{\mathrm{d}_{t^A} h(\boldsymbol{w}) + \theta \, \mathrm{d}_{t^A} h(\boldsymbol{w})}{(D_{\boldsymbol{w}} \hat{h}(\boldsymbol{w}))^2}. \tag{A.6}$$

Recall that for $v(\boldsymbol{w})$ given by

$$v(\boldsymbol{w}) = v_1 + \eta \, \hat{v}_{\frac{1}{2}} + w \, v_0 + \cdots, \tag{A.7}$$

$$B(v) = (v_1 \, b_{-1} + \hat{v}_{\frac{1}{2}} \, \beta_{-\frac{1}{2}} + v_0 \, b_0 + \cdots) + c.c., \tag{A.8}$$

where the terms replaced by the ellipsis annihilate $|D_{\mathrm{gh}}\rangle$. Expansion coefficients of $v(\boldsymbol{w})$ can be easily found by expanding eq. (A.6) around $\boldsymbol{w} = 0$. It is important to remember, though, that $h$ and $\hat{h}$ are not independent, but satisfy the superconformal constraint $D_{\boldsymbol{w}} h = \hat{h} \, D_{\boldsymbol{w}} \hat{h}$, and therefore all the expansion coefficients of $h(w)$, except for its zero mode, are determined by those of $\hat{h}$. Explicitly,

$$\begin{aligned} \hat{h}(\boldsymbol{w}) &= \theta_0 + \eta \, a + w \, \alpha + \cdots \\ h(\boldsymbol{w}) &= z_0 + \eta \, a \, \theta_0 + w \, (a^2 + \alpha \, \theta_0) + \cdots, \end{aligned} \tag{A.9}$$

where all coefficients depend on $t^A$.

In principle, we could substitute these expansions directly into eq. (A.6) and find the expansion coefficients of $v(\boldsymbol{w})$, but it is easier to find $D \, v(\boldsymbol{w})$ first. A dimension $(-1/2, 0)$ superconformal field $D \, v(\boldsymbol{w})$ is determined solely by $\hat{h}(\boldsymbol{w})$ and is given simply by

$$D_{\boldsymbol{w}} \, v(\boldsymbol{w}) = -2 \frac{\mathrm{d}_{t^A} \hat{h}(\boldsymbol{w})}{D_{\boldsymbol{w}} \hat{h}(\boldsymbol{w})}. \tag{A.10}$$

Equation (A.6) can then be used to determine the zero mode, $v_1$. In this approach, a short calculation gives the result:

$$v_1 = \frac{\mathrm{d}_{t^A} z_0 - \theta_0 \, \mathrm{d}_{t^A} \theta_0}{a^2} \tag{A.11}$$

$$\hat{v}_{\frac{1}{2}} = \frac{\mathrm{d}_{t^A} \theta_0}{a} \tag{A.12}$$

$$v_0 = \frac{\mathrm{d}_{t^A} a}{a} - \frac{\alpha \, \mathrm{d}_{t^A} \theta_0}{a^2} \tag{A.13}$$

It is natural to use complex coordinates $z_0$ and $\theta_0$ as the parameters determining $H = (h, \hat{h})$ instead of arbitrary so far $t^A$. In order to simplify notation we will drop the subscripts of $z_0$, $\theta_0$ and $d_t$ (the last one will now be $d_z$).

Using eqs. (A.11)-(A.13), one can find that $\omega_\chi$ and $\omega_D$ are given by the following expressions

$$\omega_\chi = \left(\mathrm{d} \log \frac{a}{\bar{a}} - \frac{\alpha \, \mathrm{d}\theta}{a^2} + \frac{\bar{\alpha} \, \mathrm{d}\bar{\theta}}{\bar{a}^2}\right) \omega^{0|1} \bar{\omega}^{0|1}, \tag{A.14}$$

$$\omega_D = \left(\bar{D}\left(\frac{\alpha}{a^2}\right) - D\left(\frac{\bar{\alpha}}{\bar{a}^2}\right)\right) \omega^{1|1} \bar{\omega}^{1|1}, \tag{A.15}$$

where $\omega^{0|1}$ and $\omega^{1|1}$ where introduced in section 3.1 (eqs. (3.1.9)).

It is easy to see that differential forms $\omega_D$ and $\omega_\chi$ are related by the de Rham differential

$$\omega_D = \mathbf{d}\,\omega_\chi. \tag{A.16}$$

This relation suggests that the sum of the integrals in eq. (A.1) vanishes according to the Stokes theorem. This is certainly not true for the reason that $\omega_\chi$ cannot be defined globally on $\Sigma_{g,n}$. Recall that the section $\sigma_* : \overline{\mathcal{M}} \to \hat{\mathcal{P}}$ defines the local coordinates $\boldsymbol{w}$ only up to a phase shift

$$w \to w \exp(i\,\phi) \tag{A.17}$$

$$\eta \to \eta \exp(i\,\phi/2). \tag{A.18}$$

The ghost dilaton insertion is invariant under this transformation, while

$$\omega_\chi \to \omega_\chi + \mathrm{d}\phi. \tag{A.19}$$

This phase dependence can be traced back to the fact that $|\chi\rangle$ does not belong to the semirelative BRST complex.

The string amplitude and therefore the factor in the dilaton theorem do not depend on the choice of the local coordinates. We will now use the freedom to choose the local coordinates in order to simplify calculations. Our approach is parallel to that used in [42] to analyze the dilaton theorem for the bosonic string.

Recall that one can define a metric on a superconformal surface by choosing a global, dimension $(1/2, 1/2)$ superconformal field $\Phi(\boldsymbol{z}, \bar{\boldsymbol{z}})$. The local coordinates $\boldsymbol{w}$ can be deduced from $\Phi(\boldsymbol{z}, \bar{\boldsymbol{z}})$ up-to a phase, requiring that

$$\mathrm{d}\boldsymbol{w}\mathrm{d}\bar{\boldsymbol{w}} = \Phi(\boldsymbol{z}, \bar{\boldsymbol{z}})\,\mathrm{d}\boldsymbol{z}\mathrm{d}\bar{\boldsymbol{z}}. \tag{A.20}$$

A short calculation shows that the forms $\omega_\chi$ and $\omega_D$ can be expressed in terms of $\Phi$ as follows

$$\omega_\chi = \left(\mathrm{d}\log\phi + D\log\Phi\,\mathrm{d}\theta + \bar{D}\log\overline{\Phi}\,\mathrm{d}\bar{\theta}\right)\omega^{0|1}\bar{\omega}^{0|1}, \tag{A.21}$$

$$\omega_D = D\bar{D}\log|\Phi|\,\omega^{1|1}\bar{\omega}^{1|1}, \tag{A.22}$$

where $\phi$ is the phase of the local coordinate, which is not defined by $\Phi$. Integrating over $\theta$ and $\bar{\theta}$, we obtain

$$\int \mathcal{D}(\theta,\bar{\theta})\mathcal{D}(\mathrm{d}\theta,\mathrm{d}\bar{\theta})\,\omega_\chi = \mathrm{d}\phi + \mathrm{d}z\,\partial\log\Phi_0 - \mathrm{d}\bar{z}\,\bar{\partial}\log\Phi_0, \tag{A.23}$$

$$\int \mathcal{D}(\theta,\bar{\theta})\mathcal{D}(\mathrm{d}\theta,\mathrm{d}\bar{\theta})\,\omega_D = \partial\bar{\partial}\log\Phi_0\,\mathrm{d}z\,\mathrm{d}\bar{z}, \tag{A.24}$$

where $\Phi_0 = \Phi|_{\theta=\bar{\theta}=0}$. Note that we use the pseudoform notation and therefore include $\mathcal{D}(\mathrm{d}\theta,\mathrm{d}\bar{\theta})$ in the integration measure. Transformation properties of $\Phi_0(z,\bar{z})$ allow us to use it as a metric on the reduced space of $\hat{\Sigma}_{g,n}$. The identification $\Phi_0 = \rho$ shows us that after the integral over odd variables is taken, $\omega_D$ turns into the curvature two-form

$$R^{(2)} = \partial\bar{\partial}\log\rho \tag{A.25}$$

and $\omega_\chi$ becomes the geodesic curvature

$$k = \mathrm{d}\phi + \mathrm{d}z\,\partial\log\rho - \mathrm{d}\bar{z}\,\bar{\partial}\log\rho. \tag{A.26}$$

At this point we can apply the Gauss-Bonnet theorem to obtain the result anticipated in eq. (A.1).

In our analysis above we did not explain why the phase angle which appears in the expression for the $\chi$-insertion is the same angle as in the definition of the geodesic curvature. This point was not made clear in [42], where the authors said that it was natural to fix the phase of the local coordinate at the boundary so that the tangent vector to the boundary had phase zero. In fact this choice is dictated by the way we choose the local coordinates in the vicinity of the compactification divisor, $\mathcal{F}_{g,n+1} \subset \overline{\mathcal{M}}_{g,n+1}$.

Recall that the surfaces in $\mathcal{F}_{g,n+1}$ can be obtained by gluing the standard three punctured sphere to a surface $\Sigma_{g,n}$ with $n$ punctures using $z \cdot w^{(3)} = q$, where $z$ is the local coordinate around one of the punctures on $\Sigma_{g,n}$, $w^{(i)}$, $i = 1,2,3$ are the local coordinates on the sphere and $|q| \leq 1$. The local coordinates $w^{(i)}$ on the standard sphere can be selected arbitrary but have to remain fixed including the phase for all values of $q$. Let $w$ be the global coordinate on the sphere and $w^{(1)} = w/r_1$, $w^{(2)} = (i\,w-1)/r_2$ and $w^{(3)} = r_3/w$), where $r_i$ are three sufficiently small positive real numbers. A the moduli parameter $q$ changes, this arrangement can be described in $z$ coordinate as follows. The first puncture stays at $z = 0$ and its local coordinate is $w^{(1)} = (z\,r_3)/(q\,r_1)$ while the second puncture moves within the circle of radius $1/r_3$ and its local coordinate is given by $w^{(2)} = (i\,z\,r_3/q - 1)/r_2$. On the boundary of $\mathcal{F}_{g,n+1}$, i.e. for $|q| = 1$, the local coordinate $w^{(1)}$ is remains the same up-to a phase, which allows us to make a smooth connection with the choice of local coordinates in $\mathcal{V}_{g,n+1}$, producing a smooth global section $\sigma_* : \overline{\mathcal{M}}_{g,n+1} \to \mathcal{P}_{g,n+1}$. One can easily see that the tangent vector to the $|z| = 1/r_3$ circle has phase 0 in the coordinate $w^{(2)}$.

# Acknowledgments

I would like to thank Louise Dolan for encouragement and support during the course of this work. Special thanks are due to James Stasheff for his interest and particularly for organizing a series of lectures where some preliminary results on the applications of super differential forms were presented. My understanding of all aspects of of integration theory on supermanifolds highly benefited from the discussions with Theodore Voronov. Communication with Theodore Voronov was a remarkable opportunity to learn this exiting mathematical theory "right from the source". I am grateful to Mark Doyle for drawing my attention to his works [47, 15] and providing me with a copy of his Ph.D. thesis [16].